\documentclass[12pt, a4paper]{article}

\usepackage[table]{xcolor}
\usepackage{amsmath}
\usepackage{amsfonts}
\usepackage{amssymb}
\usepackage{graphicx, rotating}
\usepackage{epstopdf}
\usepackage{epsfig}
\usepackage{subfigure}
\usepackage{latexsym}
\usepackage{graphicx}
\usepackage{wasysym}
\usepackage{color}
\usepackage{amsmath,amssymb}
\usepackage{cite}
\usepackage{slashed}
\usepackage{hyperref}
\hypersetup{colorlinks, citecolor=bluscuro, linkcolor=black, urlcolor=bluscuro}
\definecolor{rossos}{cmyk}{0,1,1,0.55}
\definecolor{bluscuro}{rgb}{0.15, 0.2, .85}
\definecolor{bluchiaro}{cmyk}{1,.3,0.,0.1}

\newcommand{\dslash}{\!\not\! \partial}
\newcommand{\eq}[1]{eq.~(\ref{#1})}

\newcommand{\nn}{\nonumber}

\newcommand{\be}{\begin{equation}}
\newcommand{\ee}{\end{equation}}
\newcommand{\bi}{\begin{itemize}}
\newcommand{\ei}{\end{itemize}}
\newcommand{\bea}{\begin{eqnarray}}
\newcommand{\eea}{\end{eqnarray}}
\newcommand{\bc}{\begin{center}}
\newcommand{\ec}{\end{center}}

\newcommand{\TeV}{\,\mathrm{TeV}}
\newcommand{\GeV}{\,\mathrm{GeV}}

\def\lra#1{\overset{\text{\scriptsize$\leftrightarrow$}}{#1}}

\def\lsim{\mathrel{\rlap{\lower3pt\hbox{\hskip0pt$\sim$}}
   \raise1pt\hbox{$<$}}}         
\def\gsim{\mathrel{\rlap{\lower4pt\hbox{\hskip1pt$\sim$}}
   \raise1pt\hbox{$>$}}}         

\begin{document}


\vspace*{-2cm}
\begin{flushright}
 CERN-TH-2016-048 \\
\vspace*{2mm}
\today
\end{flushright}

\begin{center}
\vspace*{15mm}

\vspace{0.8cm}
{\Large \bf Patterns of Strong Coupling for LHC Searches}

\vspace{0.9cm}

{\bf Da Liu$^{a,b}$, Alex Pomarol$^{c,d}$, Riccardo Rattazzi$^{b}$, Francesco Riva$^{c}$}

\vspace{.5cm}
\it {$^a$\,State Key Laboratory of Theoretical Physics, Institute of Theoretical Physics,\\ Chinese Academy of Sciences, Beijing, People's Republic of China}

\it {$^b$\,Theoretical Particle Physics Laboratory, Institute of Physics,\\ EPFL,  CH--1015 Lausanne, Switzerland}
\centerline{$^c${\it CERN, Theoretical Physics Department,  1211 Geneva 23, Switzerland}}
\centerline{$^d${\it Dept. de F\'isica and IFAE-BIST, 
Universitat Aut\`onoma de Barcelona, 08193 Bellaterra, Barcelona}}

\end{center}

\vspace*{10mm} 
\begin{abstract}\noindent\normalsize
Even though the Standard Model (SM) is weakly coupled at the Fermi scale,  a new strong dynamics involving its degrees of freedom
may conceivably lurk at slightly higher energies, in the multi TeV range. Approximate symmetries  provide a structurally robust context where, within the low energy description,
the dimensionless SM couplings are weak, 
while the new strong dynamics manifests itself exclusively through higher-derivative interactions. We present an exhaustive classification of such scenarios in the form of effective field theories, paying special attention to new classes of models where the strong dynamics involves, along with the Higgs boson,   the SM gauge bosons and/or the fermions. 
The IR softness of the new dynamics
suppresses its effects at LEP energies, but deviations are  in principle detectable at the LHC, even at energies below the threshold for production of new states.
Our construction provides the so far unique structurally robust context where to motivate several searches in Higgs physics, diboson production, or $WW$ scattering, which were so far poorly justified.
Perhaps surprisingly, the interplay between weak coupling, strong coupling and derivatives, which is controlled by symmetries, can override the naive expansion
in operator dimension, providing  instances where dimension-8 dominates dimension-6, well within the domain of validity of the low energy effective theory. This result reveals the limitations of an analysis that is both ambitiously general and restricted to dimension-6 operators.

\end{abstract}

\vspace*{3mm}

\newpage

\section{Introduction}
The primary mode of exploration of the energy frontier at 
the Large Hadron Collider (LHC)  is  through the search for  on-shell produced  new states. A secondary but  nonetheless important mode is through the study of the high-energy behavior of SM processes, as they can be affected by the presence of  off-shell heavy states. This second mode is arguably more important the stronger the new dynamics \cite{other}. The most motivated instance of such a new strong dynamics at around the weak scale is given by the
scenario of Higgs compositeness~\cite{gk,Agashe:2004rs}, which  represents one of the very few solutions of the hierarchy problem. In that scenario  the new  dynamics necessarily concerns the Higgs boson, the longitudinal components of electroweak vector bosons and possibly the third quark family.

 The direct involvement  in the strong dynamics of the other degrees of freedom, the transverse polarizations of the vector bosons in particular, is  more like an option that does not even seem well motivated both from a phenomenological and a model building perspective.
In particular, as concerns vector compositeness, the
 literature seems to lack a structurally robust scenario, based on symmetries and dynamics, which  implements it. 
It is perhaps for that reason that, while the implications of Higgs (and top) compositeness at the LHC have been widely and carefully  studied,  those of vector compositeness have not.  That is certainly a pity given the great amount of data the LHC will be harvesting on these particles. 

The main goal of this paper is to construct,  from an effective field theory (EFT) perspective,  structurally robust scenarios of strong coupling. We will classify a number of different scenarios where the SM fermions, vectors or scalars become involved in the strong dynamics, and discuss their implications. 
An interesting corollary of our analysis is that structurally robust ``accidents", due to symmetry and dynamics, can overcome the naive suppression of the energy expansion and boost the effects of dimension-$8$ operators above that of dimension-$6$ operators, and that within the domain of validity of the EFT description.

This paper is organized as follows.
In Section~\ref{SILH} we review the case of a Strongly-Interacting Light Higgs (SILH) and use it as a playground to introduce our logic and to apply it to different hypotheses, such as that of an accidentally light Higgs (ALH), detailed in Appendix~\ref{sec:ALH}. In Section~\ref{sec:dipolestrong} we present a picture for vector compositeness, where only higher derivative  interactions are associated with a strong coupling. Similar ideas are behind the discussions in Section~\ref{sec:stronglfermions}, where an approximate extended supersymmetry favors higher-derivative strong interactions involving fermions. In Section~\ref{sec:EW} we combine these ideas to depict the possible patterns of strong coupling at around the weak scale. We also leave for the Appendices detailed discussions of the Minimal Coupling (MC) assumption and of the specific dimension-8 operators that are relevant to our analysis.

\section{The SILH, its Operators and  its Power-Counting}
\label{SILH}
Before  introducing new proposals for  strongly coupled dynamics  at the weak scale, 
we  would like to recall the Strongly-Interacting Light Higgs~\cite{Giudice:2007fh} (SILH) scenario -- the reader familiar with these concepts can skip to Section~\ref{sec:dipolestrong}. 
In our view the SILH represents a  benchmark in the domain of weak-scale effective Lagrangians, and that  for various reasons: its motivation by  the hierarchy problem, its simplicity and its robustness. In particular, the SILH construction is made robust by the existence of at least one class of explicit UV completions \cite{Agashe:2004rs},  based on warped 
extra-dimensions \cite{Randall:1999ee}.  A brief review of the SILH will also allow us to introduce its operator basis,  which we shall employ in the rest of the paper, as well as to synthetically explain some points, which perhaps lend themselves to misunderstandings
(in Appendix \ref{minimalcoupling} we further  digress on the notion of MC).

The SILH relies on the basic  assumption that the  microscopic theory basically consists of two sectors. On one hand  there are  the  SM fermions and gauge bosons,  and on the other  there is an approximately conformally invariant strongly coupled sector, which we simply dub the CFT and from which the Higgs sector originates. 
The CFT is weakly coupled to the SM 
and  develops a mass-gap at around the TeV scale. 
 The Higgs doublet emerges as a light composite degree of freedom, 
and  can be naturally light, provided it  is a pseudo-Nambu-Goldstone boson (PNGB)  \cite{gk} living in a coset ${\cal G/H}$, of some approximate global symmetry ${\cal G}$ of the CFT. 
A nice property of this scenario  is that the gauge and proto-Yukawa couplings can conceivably be the only source of explicit breaking of ${\cal G}$, and consequently be fully responsible for the electroweak vacuum dynamics. With the constraint of custodial symmetry and group compactness\footnote{Group compactness follows from the request of unitarity, if one wants to interpret ${\cal G}$ as a global symmetry of the CFT. We shall come back to this point in the next section.},
 there is a unique option for the minimal coset fitting just the Higgs doublet and no other scalar: $SO(5)/SO(4)$ \cite{Agashe:2004rs}. The simplest next options would be $SO(6)/SO(5)$ ~\cite{Gripaios:2009pe, Galloway:2010bp}
 and $SO(6)/SO(4)\times SO(2)$ \cite{Mrazek:2011iu} involving respectively an additional singlet and an additional doublet. 
 
 It   is also legitimate  to entertain the possibility of a composite scalar that happens to be  light because of some unexplained accident.  As we shall discuss below and explain in more detail in Appendix~\ref{sec:ALH}, such an Accidentally Light Higgs (ALH) 
 can serve as an instructive term of comparison.  
Yet another possibility is to have a light Higgs boson arising from an accidental  global supersymmetry of the CFT  \cite{lightsusyhiggs}.  Nevertheless, we will not   discuss this last possibility here any further.

 \subsection{The One-Coupling--One-Scale Assumption}
 The simplest possible assumption for the CFT dynamics  at the TeV scale is that it  be  broadly characterized  by just one  scale $m_*$, describing the mass of the resonances,
 and one coupling $g_*$, describing their interactions.
One can certainly consider scenarios with more structure than that. 
Nonetheless it is reassuring to know that the simplest example of holographic composite Higgs model \cite{Agashe:2004rs}
obeys that simple structure, with the role of $m_*$ and $g_*$ played respectively by the   Kaluza-Klein (KK) mass  $m_{KK}$ and  the coupling arising  from 5D gauge interactions:
 \be
 m_*\sim \frac{1}{R}\equiv m_{KK}\ ,\quad\quad g_*\sim \frac{g_5}{\sqrt {\pi R}}\equiv g_5 \sqrt \frac{m_{KK}}{\pi}\,,
 \label{KKg_5}
 \ee
where $\pi R$ is the compactification length and $g_5$ the 5D gauge coupling. 
 A slightly more general situation would arise in string theory or in  generic large-$N$ gauge theories. For instance in type I string theory, the role of $m_*$ would be played by the string scale, but there would be two different couplings, for open and for closed string states,  respectively scaling  like $\sqrt g_S$ and $g_S$, where $g_S$ is the string coupling. Similarly, in the case of large-$N$ gauge theories,  $m_*$ identifies the lightest hadron mass  and  two couplings, $g_*\sim 4\pi /\sqrt N$ and $g_*'=4\pi/N$, describe  respectively the interactions of  mesons and glueballs. It would be interesting to also consider   more general situations. Nevertheless, one is easily convinced that, as long as the SM fields couple to just mesonic operators, no differences will emerge in practice. 
 
Now,   below the scale $m_*$, 
we can write an effective Lagrangian for the 
composite resonances (scalars, vectors or spinors), denoted collectively by $\Phi$, 
which, by a symmetry or just by accident, happen to be lighter than $m_*$.
This  Lagrangian
can be written as a ``loop" expansion\footnote{In the case of phenomenological models based on 5D constructions this corresponds to an expansion in loops involving the KK-modes. The same structure arises also in string theory and large-$N$ gauge theories from an expansion in topologies, of the worldsheet and of Feynman diagrams respectively.}:
\be
{\cal L}_{eff}=\frac{m_*^4}{g_*^2} L\left(\frac{\Phi}{m_*}, \frac{\partial_\mu}{m_*}\right)=
\frac{m_*^4}{g_*^2} \left \{L_0\left(\frac{\Phi}{m_*}, \frac{\partial_\mu}{m_*}\right)+\frac{g_*^2}{16\pi^2}L_1\left(\frac{\Phi}{m_*}, \frac{\partial_\mu}{m_*}\right)+\dots\right \}\,,
\label{effectivelagrangian0}
\ee
where we are   working with  non-canonically normalized fields, and
denoting by   $L_n$   generic functions.
Effective Lagrangians so structured will be at the basis of all our discussions.
From them it will be  straightforward to power count the coefficients of the higher-dimensional operators deforming the SM Lagrangian. 
 In particular, one finds that  scattering amplitudes involving $n$-quanta scale like $
{\cal A}_n\propto g_*^{n-2}$.

\subsection{Partial Compositeness}\label{sec:partComp}
In composite Higgs models  it is usually assumed that   the SM fermions and gauge bosons
are not   part of the CFT,  and  that their involvement can be dialed by the choice of  their mixings with the 
strongly coupled states\cite{Kaplan:1991dc,Agashe:2004rs}. Those mixing parameters can also be viewed as a measure of the (partial) compositeness of the corresponding   SM state.  In models based on warped extra-dimensions  these parameters codify the shape of the wave function of the SM states: 
the more the wave function is peaked towards the throat of AdS, the more composite the state is. 
This state of affairs can be neatly described in the operator language by  the use of the AdS/CFT correspondence~\cite{Maldacena:1997re}. 
In all models considered in the literature so far, the following operator mixings played a leading role
\be
{\cal L}_{mix}=\epsilon_A A_\mu J^\mu + \epsilon_\psi \psi\, {\cal O}_\psi 
+h.c.\, ,
\label{mixingJO}
\ee
where $A_\mu$ and $\psi$ indicate generically gauge fields and fermions of the SM, while $J^\mu$ and ${\cal O}_\psi$ are respectively gauge currents and  composite fermion operators of the strong sector. The coupling to $A_\mu$ just corresponds to the weak gauging of a $SU(3)\times SU(2)\times U(1)$ subgroup of the global symmetry ${\cal G}$ of the CFT.
We have dropped all internal indices to simplify notation.  To properly interpret the $\epsilon$'s, we must also specify the normalization of the fields, which amounts to specifying the strength of the amplitude with which they interpolate between the vacuum and
particle states. Canonical $A_\mu$ and $\psi$ interpolate with strength $1$ for SM one-particle states, while canonical $J^\mu$ and ${\cal O}_\psi$ interpolate  with amplitude $\sim g_*^{n-1}$
for $n$ particle states from the strong sector. 
Considering the case $n=1$,  we can thus interpret the $\epsilon$'s in eq.~(\ref{mixingJO}) as the mixing between elementary and composite quanta with corresponding quantum numbers. 
One is easily convinced that the microscopic mixing Lagrangian in eq.~(\ref{mixingJO}) corresponds  to an  effective low-energy Lagrangian of the form
\be
{\cal L}_{eff}=\frac{1}{g_*^2} \left \{m_*^4L\left(\frac{\Phi}{m_*}, \frac{D_\mu}{m_*},\frac{\epsilon_A \hat F^i_{\mu\nu}}{m_*^2},
\frac{\epsilon_\psi \hat \psi}{m_*^{3/2}}\right)-\frac{1}{4}(\hat F^i_{\mu\nu})^2+i\bar {\hat \psi} \gamma^\mu D_\mu\hat \psi\right \}\,,
\label{lmess}
\ee
where, 
\bea
D_\mu&\equiv& \partial_\mu+i\epsilon_A T_i\hat A^i_\mu\,, \label{coder}\\
\hat F_{\mu\nu}^i&\equiv& \partial_\mu \hat A^i_\nu-\partial_\nu \hat A^i_\mu -\epsilon_A f^{ijk}\hat A_\mu^j\hat A_\nu^k \,,
\label{Fmunu}
\eea
with $T_i$ and  $f^{ijk}$  respectively the generators and  structure constants of the gauge group. 

Several comments on the above equation are in order.
Given $[D_\mu,D_\nu]=\epsilon_A \hat F_{\mu\nu}$, the exhibited explicit dependence of $L$ on the field-strength is redundant: we exhibit it only in order to better stress the difference with other scenarios we shall discuss later. Furthermore, as it was the case for  \eq{effectivelagrangian0}, and as
 we shall better discuss in the next section,  $L$ is given by of a loop expansion.
 Notice also the non-canonical normalization of all the fields, including the elementary ones: $ \hat A_\mu \equiv g_* A_\mu$ and $\hat \psi\equiv g_*\psi$. When $\epsilon_{A,\psi}\to 0$, the SM fields $A_\mu$ and $\psi$ decouple. Accordingly,
the SM gauge couplings are given by $g_A\equiv \epsilon_A g_*$:  the universality of gauge interactions fixes the mixing $\epsilon_A$. We stress 
that if it weren't for the  explicitly exhibited elementary field kinetic terms, the $\epsilon$'s could be absorbed into the definition of $A_\mu$ and $\psi$, and these would be  coupled as strongly as the $\Phi$'s.
 The explicit kinetic term is precisely what  characterizes $A_\mu$ and $\psi$ as elementary fields.

Eq.~(\ref{lmess}) gives rise to contributions to  the scattering  amplitude among   $n_e$ elementary and $n_c$ composite states satisfying the scaling
\be
{\cal A}_{n_e,n_c} \propto \epsilon^{n_e} g_*^{n_e+n_c-2}\, ,
\label{simplepower}
\ee
where  $\epsilon$ indicates collectively $\epsilon_A$ and $\epsilon_\psi$, and  we omitted the energy dependence.
As already said, the ordinary  gauge couplings are $g_A\equiv  \epsilon_A g_*$.  On the other hand, in the fermion sector there is more freedom.  As the Higgs
multiplet is issued from the strong dynamics,   the Yukawa coupling will have the form (with an obvious notation)
\be
y_\psi\sim \epsilon_{\psi_L}\epsilon_{\psi_R} g_*\, ,
\ee
so that only the product $\epsilon_{\psi_L}\epsilon_{\psi_R}$ is fixed. 
In principle, it is possible, and sometimes phenomenologically convenient, to have either $\epsilon_{\psi_L}$ or $\epsilon_{\psi_R}$ to be large. In particular, models with $\epsilon_{t_R}\sim 1$ are especially attractive \cite{Mrazek:2011iu}.
Notice that this  situation can arise naturally if the operator ${\cal O}_{t_R}$ mixing to $t_R$,   has dimension
 satisfying  $3/2<{\rm dim}[{\cal O}_{t_R}]<5/2$, such that the interaction of $t_R$ with the CFT is relevant and becomes strong at low-energies, turning {\it {de facto}} $t_R$ into a  strongly-interacting composite fermion.

\subsection{The Effective Lagrangian at the Electroweak Scale}
Putting together the concepts presented above, it is  straightforward to derive the general structure of the effective Lagrangian for a strongly-interacting light Higgs.  This corresponds to  reduce the general scenario of the previous section 
 to the simplest case in which the set of light states from the CFT  contains  just the Higgs doublet.
In this case, the $L$ in \eq{lmess} is given by 
\bea
L&=&\sum_n \left (\frac{g_*^2}{16\pi^2}\right )^n L_n\,, \label{loopexp}\\
L_n&\equiv&  L_n\left(\frac{\hat H}{m_*},\frac{D_\mu}{m_*},\frac{\epsilon_A\hat F^i_{\mu\nu}}{m_*},
\frac{\epsilon_\psi\hat \psi}{m_*^{3/2}}\right )\,,
\eea
with  $\hat H$ indicating the non-canonically normalized Higgs doublet field.
In the leading term $L_0$  the SM  (fermion and gauge) fields appear only  as external spectators. 
The one-loop term $L_1$ can instead be written as the sum of a pure CFT contributions, $L_1^{(0)}$, and   contributions 
arising from  the  exchange of one virtual SM state:
\be\label{LNL}
L_1=L_1^{(0)}+\epsilon^2L_1^{(1)}\,.
\ee
The above structure generalizes in the obvious way at the higher loop order.

 In the simplest and well motivated situation where the Higgs $H$ spans the  $SO(5)/SO(4)$ coset, the above Lagrangian will respect the symmetry and the selection rules it imposes. The $H$ dependence will arise via the Goldstone matrix $U=e^{i \Pi}$ and will be determined by the CCWZ construction \cite{Coleman:1969sm},
 while the dependence on the symmetry-breaking mixings  $\epsilon$'s will be constrained by the $SO(5)$ selection rules (see ref. \cite{Panico:2015jxa} for a thorough discussion).
 
Within our assumptions, the estimate of the operator coefficients in the effective Lagrangian  is a simple matter of power counting. 
Using the definition of the dimension-6 operators ${\cal O}_i$ in table \ref{operatorsdim6}, we illustrate the leading results  for  three different scenarios in table \ref{tab:Silh}. 
The   coefficients $c_i$ are defined by 
\be
{\cal L}_6=\frac{1}{m^2_*}\sum_i\,  c_i\,{\cal O}_i\,.
\ee
The first scenario in table \ref{tab:Silh} corresponds to an ALH that happens to be lighter than the other resonances just because of some accidental cancellation, and not because of a symmetry --see Appendix \ref{sec:ALH}. The second scenario we consider
is that of a general PNGB strongly-interacting light Higgs (GSILH), defined by the most
general $L$ satisfying the $SO(5)$ selection rules. The third scenario is the slightly more specific case of
the  SILH considered in \cite{Giudice:2007fh},  where $L_0$ is not completely generic  because of  restricted properties of the dynamics at the scale $m_*$. This third class describes, for instance, Little Higgs models and Holographic composite Higgs models.

 \begin{table}[t]
\begin{center}
\begin{tabular}{|l|}\hline
${\cal O}_H=\frac{1}{2}(\partial^\mu |H|^2)^2$\\
${\cal O}_T=\frac{1}{2}\left (H^\dagger {\lra{D}_\mu} H\right)^2$\\
${\cal O}_6= |H|^6$
  \\    \hline  \hline  
${\cal O}_W=\frac{i}{2}\left( H^\dagger  \sigma^a \lra {D^\mu} H \right )D^\nu  W_{\mu \nu}^a$\\
${\cal O}_B=\frac{i}{2}\left( H^\dagger  \lra {D^\mu} H \right )\partial^\nu  B_{\mu \nu}$\\
\hline
${\cal O}_{HW}=i (D^\mu H)^\dagger\sigma^a(D^\nu H)W^a_{\mu\nu}$\\
${\cal O}_{HB}=i (D^\mu H)^\dagger(D^\nu H)B_{\mu\nu}$\\
\hline
${\cal O}_{BB}=
 |H|^2 B_{\mu\nu}B^{\mu\nu}$\\
${\cal O}_{GG}=
|H|^2 G_{\mu\nu}^A G^{A\mu\nu}$\\
\hline
 \end{tabular}\hspace{10mm}
 \begin{tabular}{|l|}\hline
${\cal O}_{y_\psi}   = |H|^2    \bar \psi_L H \psi_R$\\
\hline\hline
${\cal O}_{2B}=-\frac{1}{2}(\partial_\rho B_{\mu\nu})^2$\\
${\cal O}_{2W}=-\frac{1}{2}(D_\rho W_{\mu\nu}^a)^2$\\
${\cal O}_{2G}=-\frac{1}{2}(D_\rho G_{\mu\nu}^A)^2$\\
\hline
${\cal O}_{3W}= \frac{1}{3!} \epsilon_{abc}W^{a\, \nu}_{\mu}W^{b}_{\nu\rho}W^{c\, \rho\mu}$\\
${\cal O}_{3G}= \frac{1}{3!} f_{ABC}G^{A\, \nu}_{\mu}G^{B}_{\nu\rho}G^{C\, \rho\mu}$\\
\hline\hline
${\cal O}^\psi_{L,R} =
(i H^\dagger {\lra { D_\mu}} H)( \bar \psi_{L,R}\gamma^\mu \psi_{L,R})$ \\ 
${\cal O}^{(3)\, \psi}_{L}=(i H^\dagger \sigma^a {\lra { D_\mu}} H)( \bar \psi_L\sigma^a\gamma^\mu \psi_L)$\\
\hline\hline
${\cal O}_{4\psi}=\bar\psi\gamma_\mu\psi\bar\psi\gamma^\mu\psi$
 \\\hline
 \end{tabular}
\end{center}
\caption{\emph{Dimension-6 operators used in our analysis. Notice that our normalization differs from previous literature.}}\label{operatorsdim6}
\end{table}
\begin{table}[t]
\renewcommand{\arraystretch}{1.3}
\begin{center}
\begin{tabular}{|c|c|c|c|c|c|c|c|c|c|c|}
\hline 
&$|H|^2$&$|H|^4$ & ${\cal O}_{H}$&${\cal O}_6$& ${\cal O}_{V}$&${\cal O}_{2V}$ & ${\cal O}_{3V}$ & ${\cal O}_{HV}$  & ${\cal O}_{VV}$&  ${\cal O}_{y_\psi}$  \\
\hline
ALH & $m_*^2$ & $g_*^2$&$g_*^2$& $g_*^4$ &${g_V}$&$\frac{g_V^{2}}{g_*^2}$& 
$\frac{g_V^2}{g_*^2}g_V$& $g_V$  &   $g_V^{2}$ & $y_\psi g^2_*$\\
GSILH & $\frac{y_t^2}{16\pi^2}m_*^2$ & $\frac{y_t^2}{16\pi^2} g_*^2$ & $g_*^2$&$\frac{ y_t^2}{16\pi^2}g_*^4$&$g_V$&$\frac{g_V^{2}}{g_*^2}$& 
$\frac{g_V^2}{g_*^2}g_V$&
$g_V$  &   $\frac{y_t^2}{16\pi^2}g^2_V$ & $y_\psi g^2_*$\\
SILH & $\frac{y_t^2}{16\pi^2}m_*^2$ & $\frac{y_t^2}{16\pi^2}g_*^2$ & $g_*^2$&
$\frac{y_t^2}{16\pi^2}g_*^4$&$g_V$&$\frac{g_V^{2}}{g_*^2}$& 
$\frac{g^2_V}{16\pi^2} g_V$& $\frac{g^2_*}{16\pi^2}g_V$
  &   $\frac{y_t^2}{16\pi^2}g^2_V$& $y_\psi g^2_*$\\
\hline
\end{tabular}
\end{center}
\caption{\emph{Estimated coefficients ($c_i$) of different  operators appearing in the effective Lagrangian for a 
strongly interacting Higgs, under different hypotheses: an accidentally small electroweak scale and accidentally light  Higgs (ALH), a general SILH (GSILH) scenario, and  the proper SILH  of \cite{Giudice:2007fh} where   the additional assumption of MC is considered (see Appendix \ref{minimalcoupling}). The subscript $V$ can denote  $W,B,G$ according to  the basis defined in table~\ref{operatorsdim6}. 
For the ALH scenario the entries in the first two columns emphasize the need for tuning, w.r.t. the NDA estimate
 (see Appendix~\ref{sec:ALH}).}}
\label{tab:Silh}
\end{table}%

 A few explanations of the results of table \ref{tab:Silh}  are in order. 
  First of all,  we should give a motivation  for our choice of operators.
   Our choice singles out ${\cal O}_{W,B}$ and ${\cal O}_{2W,2B}$ as the only operators involving  vectors that can be generated at tree level by the exchange of massive vectors in a  renormalizable theory. Now, as it turns out, 
  the Little Higgs models and holographic Higgs models, in their simplest incarnations, at the scale $m_*$ are described to a rather good approximation by renormalizable Lagrangians. That property is essentially a corollary of the mechanism of collective breaking enforced in these models: the scale of the resonances $m_*$ is parametrically smaller than the genuine cut-off $\Lambda$ of the theory, at which no weakly coupled description is tenable. 
In such models, apart from ${\cal O}_{W,B}$ and ${\cal O}_{2W,2B}$, all  operators involving vectors can only be generated via loops. Such a structure was dubbed in Ref.~\cite{Giudice:2007fh}, for somewhat obvious reasons, Minimal Coupling (MC). In   Appendix \ref{minimalcoupling} we give a detailed discussion of MC in the context of extra-dimensional models.
   A second important property of our choice of basis is that we only need two operators of the form $H^\dagger H F_{\mu\nu}F^{\mu\nu}$, which we choose to be ${\cal O}_{GG}$
 and ${\cal O}_{BB}$. Now, it turns out that in the case where the Higgs is a PNGB, the coefficient of these two operators is subject to the same selection rules, and thus to the same protection, that control the Higgs potential. The reason is quite simple, and can be understood even without delving into the details of  $SO(5)$ and of its breaking: if  we were to turn-off all interactions of the SM apart from color and electric charge\footnote{Notice that this would amount to just gauging $U(1)_{\rm EM}\subset SU(2)_L\times U(1)_Y$.},
 the neutral  Higgs $h$ would remain an exact Goldstone boson. 
 Therefore, in such a limit, $h$ would have to appear always coupled derivatively.
 Since  ${\cal O}_{GG}$ and ${\cal O}_{BB}$ contain  non-derivative Higgs interactions
  in the presence of gluonic and photonic backgrounds, their coefficients  should be   suppressed. 
  In practice that means that in the case of a PNGB Higgs,  ${\cal O}_{GG}$ and ${\cal O}_{BB}$ are absent in $L_0$ and appear first in $\epsilon^2L_1^{(1)}$ in \eq{LNL}, together with the leading contribution to the Higgs potential. 
Both for ${\cal O}_{GG},\,{\cal O}_{BB}$ and for the potential the leading contributions  come 
  from the top sector at 1-loop. These effects carry some model dependence, associated with the choice of  the $SO(5)$ quantum numbers for the CFT operators that mix with the  top  as in eq.~(\ref{mixingJO}). 
   The estimates in table~\ref{tab:Silh} correspond to the scenario where  the  need for tuning in the Higgs potential 
 is minimized. That is the scenario where either $\epsilon_{t_L}\sim y_t/g_*$, $\epsilon_{t_R}\sim 1$ or 
 $\epsilon_{t_L}\sim 1$, $\epsilon_{t_R}\sim y_t/g_*$ \cite{Panico:2015jxa}, in which case the coefficients of ${\cal O}_{GG},\,{\cal O}_{BB}$ feature an extra suppression
$\propto {y_t^2/g_*^2}$ \cite{Giudice:2007fh}.

\section{Effective Theories of Strong Multipolar Interactions} \label{sec:dipolestrong}
We are interested in exploring  patterns of strong coupling  beyond the SILH  scenario described above.
In particular, we would like to  consider the case in which the SM vector bosons also arise from some strong dynamics at the TeV.  At first sight, however, that seems difficult to reconcile 
 with the  observed weakness of the SM gauge interactions (see, for instance, ref.~\cite{Csaki:2011xn} to get a better taste of the difficulties).
Indeed,  one could  conceive a situation where the SM vector bosons are composite in the multi-TeV range and the weakness of their effective coupling results from the large number of degrees of freedom in the underlying dynamics, like it happens 
 in large-$N$ gauge theories.
However,  in that situation the coupling of the SM vector bosons to the heavy  resonances would also be suppressed,
so that the new dynamics would not appear as genuinely strong, or, more precisely,  it would not appear stronger than the SM dynamics. 
In order to counter that, the underlying dynamics should be more structured and involve at least two couplings, a weak one $g$, describing the low-energy strength of the gauge interactions, and a stronger one $g_*$, describing the interactions of gauge bosons with massive resonances. It would also be desirable to have such state of affairs be made structurally robust by symmetry. That could in principle provide, even in the absence of an explicit UV complete realization of the scenario, a set of rules to consistently power count the coefficients in the effective Lagrangian, as in the SILH case.

A technically natural situation where the vector bosons are involved in a dynamics with coupling $g_*> g$ just above the electroweak scale can be pictured as follows. Notice first of all that the gauge coupling $g$, the one appearing in the covariant derivative, controls the {\it {monopole}} charge ({\it {electric}}, {\it{chromo-electric}} etc. for the various gauge factors of the SM) of the light degrees of freedom. The effective Lagrangian, however, also describes all {\it {higher-multipole}} interactions, which are unavoidably associated with the existence of structure at the fundamental scale~$m_*$. It is then intuitive that there could exist situations in which the resonances (at least the light ones) have small charges, controlled by $g$, but  large multipoles, controlled by  $g_*$. The simplest limiting situation 
would be a strong dynamics producing a composite photon
$A_\mu$ without any charged light degrees of freedom.\footnote{We do not need to worry about  the weak gravity conjecture\cite{ArkaniHamed:2006dz}: in this limiting case, conflict is avoided provided heavy
charged resonances at the scale $m_*$ exist, while  in a more realistic case the light states have small but non-zero charges.} As a matter of fact  the  effective theory is similar to the  one resulting in the SM
below the electron mass scale, with the photon and the neutrinos as the only remaining degrees of freedom.

The effective theory for a photon $A_\mu$ with multipolar interactions
 depends only on  the field-strength $F_{\mu\nu}$ and on its derivatives:
\be
{\cal L}_{eff}=\frac{m_*^4}{g_*^2} L\left(\frac{\hat F_{\mu\nu}}{m_*^2},\frac{\partial_\mu}{m_*},\frac{\hat \Phi}{m_*} \right)\,,
\label{effectivelagrangian}
\ee
where $\hat \Phi$ denotes other, neutral,  degrees of freedom of the theory (again hatted fields are   non-canonically normalized). 
 While  $A_\mu$   behaves like a free field at low-energies,  its interactions grow with energy. 
  For instance, the amplitude for light by light scattering is proportional to $g_*^2 E^4/m_*^4$, and reaches a strength $g_*^2$ at $E\sim m_*$. Starting from such an obviously technically natural situation, one can imagine deforming the effective theory by endowing the light resonances with small charges. That amounts to deforming the ordinary derivatives into covariant derivatives:
\be
\partial_\mu \Phi\,\to\, (\partial_\mu +i\epsilon q_\Phi A_\mu)\Phi\,,
\ee
where $q_\Phi \sim O(1)$ while  $\epsilon \equiv {g}/{g_*}\ll 1$. Again, as the covariant derivatives are not renormalized, the smallness of $\epsilon$ is technically natural. 
It realizes the intuitive situation where the light composites have small charges but large multipoles.

It is interesting to generalize the above situation to the case of $N_A$ 
 vectors $A^i_\mu$  of a non-abelian gauge group $\cal G$. 
Neglecting for simplicity matter fields, the strong dynamics  must have the following features: 
\begin{itemize}
\item There  must  exist  $N_A$ composite photons  associated with the gauge group $U(1)^{N_A}$.
\item  The  non-abelian symmetry $\cal G$ must be a global symmetry of the strong sector   
under which the photons transform in the adjoint representation of dimension $N_A$.
\end{itemize} 
The $U(1)^{N_A}$ gauge symmetry guarantees that we have $N_A$ massless vectors, each coming with  two helicities, while the global symmetry  $\cal G$ is needed to   render the gauging of  $\cal G$  a small deformation, as we will describe below.
Notice that the $U(1)^{N_A}$ generators sit in the adjoint of $\cal G$, so that the symmetry group is actually the semidirect product $[{\cal G}]_{global}\rtimes [U(1)^{N_A}]_{local}$.
Again, as before, all fields $\Phi$ must be neutral under the gauge group, so that all interactions, in principle strong, are dominated by irrelevant higher-derivative interactions. 
All this can be encoded in a general one-coupling--one-scale effective Lagrangian as in eq.~(\ref{effectivelagrangian}). Now, starting from this theory, we can consider the smooth deformation of its symmetry according to
\be
[{\cal G}]_{global}\rtimes [U(1)^{N_A}]_{local}\,\to \, [{\cal G}]_{local}\label{deform}\,.
\ee
Notice that although the gauge group is modified, the deformation is still smooth. In particular, the number of local generators is unaffected, so that the number of  degrees of freedom is unchanged.
As in the abelian case, the deformation will simply amount to replacing   derivatives 
 and field-strengths  with their covariant form in    \eq{coder} and  \eq{Fmunu}, 
with  $\epsilon_A\to \epsilon$.\footnote{Notice that we use the normalization implied by eq.~(\ref{effectivelagrangian}): the undeformed kinetic term is~$-1/{4g_*^2} F_{\mu\nu}F^{\mu\nu}$.}
  The corresponding effective  Lagrangian will now have  the form
\be
{\cal L}_{eff}=\frac{m_*^4}{g_*^2} L\left(\frac{\hat F^i_{\mu\nu}}{m_*^2},\frac{D_\mu}{m_*}\right)\,,
\ee
where, contrary to \eq{lmess},   the field-strength $\hat F^i_{\mu\nu}$ appears without a suppression factor.
Higher-derivative interactions are thus  controlled by the strong coupling $g_*$ and by the scale $m_*$, 
while IR physics is controlled by the (classically) dimensionless weak coupling $g= \epsilon g_*$. The structure of our Lagrangian is natural, in the sense that it is the most general compatible with the symmetries, both exact and approximate.  In particular, in the limit
$\epsilon=0$, where the deformation is turned off, our Lagrangian is the most general one respecting $[{\cal G}]_{global}\rtimes [U(1)^{N_A}]_{local}$ and described  by a scale $m_*$ and a strong coupling $g_*$, which in principle could be as large as $\sim 4\pi$. Once $\epsilon$ is turned on, we have the most general Lagrangian respecting an exact $[{\cal G}]_{local}$ and an approximate $[{\cal G}]_{global}\rtimes [U(1)^{N_A}]_{local}$.
 The important difference with respect to the usual discussion of naturalness of small parameters is that in our case  $\epsilon$ deforms the symmetry into a group which is not a subgroup of the original one. In other words, $\epsilon=0$ is not a point of enhanced symmetry, but  a point of deformed symmetry. The consequences for naturalness are however the same: symmetry singles out $\epsilon=0$  as a stable point.

  As a matter of fact the situation encountered in our construction is fully analogous to the one encountered when going from the Galilei to the Poincar\`e group: the number of generators is unchanged but the group structure (and the corresponding Lie algebra) is modified. In that case the parameter playing the role of $\epsilon$ is represented by the inverse of the speed of light $1/c$: when $c\to\infty$, with everything else fixed, the Poincar\`e group is deformed into the Galilei group. According to that limiting procedure, the Galilei group is said to be an Inonu-Wigner (IW) contraction of the Poincar\`e group~\cite{IW}.
It is here worth recalling that the IW contraction of a Lie group ${\cal G}$ with respect to a subgroup  ${\cal H}$    corresponds to a group with the same Lie algebra as  $\cal G$, except for the commutators among the generators
in ${\cal G}/{\cal H}$, which are {\it contracted} to zero. The contracted group is thus simply ${\cal H}\rtimes [{\cal G}/{\cal H}]_{abelian}$. The IW contraction precisely describes the relation between the two symmetry
groups in \eq{deform}: $[{\cal G}]_{global}\rtimes [U(1)^{N_A}]_{local}$ is the contraction of the gauge group $[{\cal G}]_{local}$ with respect to its global subgroup $[{\cal G}]_{global}$.  In practice all local generators are abelianized.

In Section \ref{sec:EW} we shall apply the  above construction to the gauge interactions of the SM and study its compatibility with cases where the Higgs and/or some of the fermions participate in the new strong dynamics. 
In the next section we will instead add  some remarks about the possibility of realizing the above scenario using the mechanism of partial compositeness.

\subsection{Strong Multipolar  Interactions and Partial Compositeness}\label{partcomp}

Following the idea of partial compositeness, one could   generalize   \eq{mixingJO}
to incorporate the coupling of the field-strength  $F_{\mu\nu}$ to the strong sector
by adding  the term
\be
\Delta {\cal L}_{mix}=\epsilon_F F_{\mu\nu} {\cal O}^{\mu\nu}\label{mixingF}\,,
\ee
where ${\cal O}^{\mu\nu}$ is some composite anti-symmetric two index tensor, normalized like in sect. \ref{sec:partComp}. The presence of this extra mixing term can be accounted for by modifying $L$ in \eq{lmess} according to
\be
L\to L\left(\frac{\hat\Phi}{m_*}, \frac{D_\mu}{m_*}, \frac{\epsilon_F \hat F_{\mu\nu}}{m_*^2},\frac{\epsilon_\psi\hat  \psi}{m_*^{3/2}}\right)\,.
\ee
In the effective theory below $m_*$, there are now two sources of couplings for $A_\mu$. 
In the presence of light charged states $\epsilon_A$ (see eq.~(\ref{mixingJO}))
 leads to   the standard  gauge coupling $g=\epsilon_A g_*$.
On the other hand, there are also higher-derivative  interactions generated by $\epsilon_F$. For such terms, the addition 
of a vector leg will cost an effective coupling 
\be
g_{eff}\sim \epsilon_F g_* \frac{E}{m_*}\,.
\ee
This is certainly suppressed at low-energies, but it grows with energy, and, for maximal mixing $\epsilon_F=O(1)$,  it  becomes  $O(g_*)$ at   $E\sim m_*$. Indeed, the limiting case $\epsilon_F=O(1)$ schematically describes the scenario of strongly-coupled dipole interactions we outlined in the previous section. Furthermore,   this  operator mixing picture makes also clear
that there is no contradiction in the choice $\epsilon_A\ll \epsilon_F\sim 1$: quantum effects from $\epsilon_F$ always involve at least one derivative acting on $A_\mu$, that is to say they represent multipole interactions that  do not affect the $A_\mu J^\mu$ term.

It must however  be remarked that the terms in eq.~(\ref{mixingJO}) are crucially distinct from all other possible operator mixings, including eq.(\ref{mixingF}): 
in theories where there exists a separation of scales and the strong sector (possibly harboring the Higgs sector) is approximately conformally invariant above the weak scale, eq.~(\ref{mixingJO}) represents all the possible relevant or marginal mixing terms.\footnote{In models in which the Higgs is elementary we could also have a marginal/relevant coupling  to the strong sector as discussed, for example, in Ref.~\cite{Samuel:1990dq}.}
 The gauge coupling is obviously marginally relevant or irrelevant as the conserved vector current is constrained by unitarity to have exactly dimension $3$ in the unperturbed CFT limit. As for  fermions, when the dimension  of ${\cal O}_\psi$ ranges from its minimal value of $3/2$, as dictated by unitarity, and $5/2$, the couplings range from relevant to marginal. 
On the other hand, if we interpret   ${\cal O}^{\mu\nu}$ in eq.~(\ref{mixingF}) as belonging to a  CFT, then the theory of unitary representation of the conformal group constrains its dimensionality to be $\geq 2$. Therefore the interaction is always irrelevant, apart for the limiting case where the dimension of  ${\cal O}^{\mu\nu}$ is exactly equal to 2. Nevertheless, in this limit  ${\cal O}^{\mu\nu}$ must correspond to a free field \cite{Ferrara:1974pt}. Indeed, it must be the field-strength of another free massless gauge field. Anyway it  does not describe states in the putative strongly interacting sector. For instance, the correlators	 of ${\cal O}^{\mu\nu}$ trivially factorize into disconnected 2-point functions and thus do not mediate any scattering among the SM vectors that are coupled to it. 
We thus conclude that it is not possible  to construct the scenarios described in the previous section  using the idea of partial compositeness in   models based on  CFTs.  By the AdS/CFT correspondence, this also tells us the impossibility
to construct these scenarios  using   warped extra-dimensions. 
In other words, there seems to be no way  to obtain the scenario of strong multipolar gauge interactions in a system which we can both  easily control, via partial compositeness,
and extrapolate to high-energies.  On the other hand, we cannot exclude that such scenario might arise 
in a QFT where the vectors are fully composite, but which cannot be continuously deformed to a QFT where they are elementary.

To conclude this discussion, and to make sure we are not missing any opportunity, it is worth commenting on all other possible mixings between the SM vectors and fermions and a CFT. Notice
 that the  terms in eqs.~(\ref{mixingJO},\ref{mixingF}) exhaust all the  possibilities for the mixing of a single SM field with a CFT primary operator. In that sense they are leading in the derivative expansion: 
any other operators  involving one fermion or one gauge boson of the SM  will have more derivatives acting on it,
and thus represent less relevant mixing to descendants of the CFT operators in  eqs.~(\ref{mixingJO},\ref{mixingF}).
One could then  consider interactions involving  SM composite operators. But also in that case one finds no options compatible with naturalness. Indeed one is quickly convinced that
 the resulting   interactions are always strictly irrelevant apart for the limiting  case where SM fermion bilinears are coupled to a CFT scalar  with dimension 1. However by unitarity this scalar field must be an approximately free field so that  its mass term will suffer from a hierarchy problem.
  This state of things encapsulates the essence of the Flavor problem of (conformal)-technicolor ~\cite{Rattazzi:2008pe}. Amusingly, the obstruction to the  construction of strong mutipolar gauge  interactions using partial compositeness and the difficulties of flavor, arise for a similar reason: the absence of relevant operators.

\subsection{Fermions as Composite Pseudo-Goldstini}\label{sec:stronglfermions}

The results of the previous section, and the known properties of NG-bosons, imply we can think of strongly coupled scenarios
where the SM vectors and the Higgs are composites of a new strong dynamics which manifests itself only through higher derivative (multipolar) interactions. In this section we want to characterize scenarios where 
the  fermions, as well, are composites with higher-derivative strong interactions. By the discussion in the previous section, and similarly to the case of a derivatively coupled scalar or vector, 
we cannot implement this scenario in a CFT using partial compositeness, as the corresponding interactions would always be irrelevant. Therefore derivatively coupled  fermions must arise directly as composites from the strong sector and cannot be deformed into elementary states.
    
Given a fermion field $\psi$, the simplest thinkable symmetry enforcing purely derivative interactions is  just given by the transformation $\psi\to \psi+\xi$. Of course the Lie parameter $\xi$ must be a Grassmann number, but notice this is not a supersymmetry, as spacetime
coordinates are not affected. Now, aside the kinetic term $i\bar\psi  \dslash \psi$, which happens to be invariant up to a total derivative, higher order invariants must be functions of   $\partial_\mu\psi$. The lowest order interaction has thus the schematic form $(\partial \psi)^4$ and arises at dimension 10. The resulting scattering amplitudes are extremely soft, behaving at low energy like $s^3\propto E^6$. However, very much like in the case of bosons,  under reasonable assumptions on the UV behaviour of the cross-section, unitarity and analyticity bound the low energy amplitude to be {\it {not softer}} than $s^2$ \cite{BellazziniTorre}. It thus seems the pure shift symmetry is disfavored by basic principles. The same conclusion of course applies to the case of multiple fermions $\psi_i$, each with its own shift symmetry $\psi_i\to \psi_i+\xi_i$.

A more plausible  alternative scenario is to consider a non-linearly realized supersymmetry which, sticking to a single Dirac fermion $\psi$ (identified with the ${\cal N}=1$ Goldstino), acts as \cite{Wess:1992cp}
\begin{equation}\label{goldstinoshift}
\delta \psi = \xi +\frac{i}{2F^2}   \partial_\mu \psi(\bar\psi\gamma^\mu {\xi}-\bar\xi\gamma^\mu \psi)\,,
\end{equation}
with $F$ the Goldstino decay constant. In a one-coupling--one-scale scenario one has the scaling $F\sim m_*^2/g_*$, which generalizes the relation
$f\sim m_*/g_*$ for the decay constant of a Goldstone boson.
The effective operators describing the coupling of a Goldstino  to itself and to other massless light  fields (such as a complex scalar $\phi$, a field-strength $F_{\mu\nu}$ or a generic matter fermion $\psi_q$) start at dimension 8, and always involve some extra derivative acting on the Goldstino field. At the lowest dimension, by use of the equations of motion, these reduce to\cite{goldstino,
Komargodski:2009rz}
\begin{gather}\label{goldststructuresN1}
\frac{i}{F^2}  \bar{\psi} (\gamma^\mu \partial^\nu+\gamma^\nu \partial^\mu) \psi F_{\mu\rho}F_\nu^{\,\,\rho}\, ,\quad \quad
\frac{i}{F^2} \partial_\mu \phi^\dagger \partial_\nu \phi\,\bar \psi(\gamma^\mu\partial^\nu+ \gamma^\nu\partial^\mu) \psi\,, \\
\frac{1}{F^2}\bar \psi^2\partial^2_\mu\psi^2\,,\quad\quad
\frac{1}{F^2}\partial_\nu \bar \psi\gamma^\mu\psi\,\, \bar \psi_q \gamma_\mu\partial^\nu\psi_q\,,\quad\quad
\frac{1}{F^2}\partial_\nu \bar \psi_q\gamma^\mu\psi \,\,\bar \psi \gamma_\mu\partial^\nu\psi_q\, .\label{goldststructuresN2}
\end{gather}
The idea here is to consider  the more general case of ${\cal N}>1$ supersymmetries, where ${\cal N}$ of the SM fermions are identified with pseudo-Goldstini from some strong dynamics (virtually all of them for ${\cal N}=45$). That would be a generalization of the Volkov-Akulov model for the  Goldstino-neutrino \cite{Volkov:1973ix}. Notice that having ${\cal N}>8$ does not force us to include massless fields of high spin: since supersymmetry is non-linearly realized, multiplets do not have to be complete. While the strong dynamics is characterized by effective interactions generalizing eq.~(\ref{goldststructuresN2}), the SM gauge and Yukawa interactions necessarily break supersymmetry, as they correspond to non-derivative interactions for the pseudo-Goldstini. However, since these other interactions are weak, the corresponding breaking can be treated as a small perturbation of the strong dynamics controlled by the more sizeable coupling $g_*$.
As a matter of fact, the situation where just the fermions, not the vectors nor the Higgs, belong to the new dynamics was already  considered long ago in an interesting paper by Bardeen and Visnjic \cite{Bardeen:1981df}. In the case of maximal  $R$-symmetry, $SU({\cal N})$,
the resulting interactions  at dimension 8 are
\begin{gather}
F^2 \det\left[\delta_{\mu\nu}+\frac{i}{2F^2}(\partial_\mu \bar \psi^a\gamma_\nu \psi^a-\bar \psi^a\gamma_\nu \partial_\mu \psi^a)\right]=i\bar \psi^a\gamma^\mu\partial_\mu \psi^a+ \frac{1}{8F^2}
\Big[(\partial_\mu \bar \psi^a\gamma^\mu \psi^a)^2+(\bar \psi^a\dslash \psi^a)^2\nn\\
-2\bar\psi^a\dslash\psi^a\partial_\nu \psi^b\gamma^\nu\bar \psi^b
-\bar\psi^a\gamma^\mu\partial_\nu \psi^a \bar\psi^b\gamma^\nu\partial_\mu \psi^b 
-\partial_\nu \bar\psi^a\gamma^\mu \psi^a \partial_\mu \bar\psi^b\gamma^\nu \psi^b
+2\bar\psi^a\gamma^\mu\partial_\nu \psi^a\partial_\mu \bar\psi^b\gamma^\nu\psi^b\Big]+\cdots\label{eq:bardeen}
\end{gather}
where  $a,b=1,...,{\cal N}$ represent the flavor indices. 
 
In principle there is no obstruction to generalize this to the case where the $R$-symmetry reduces to  a subgroup of $SU({\cal N})$ and to the case where the vectors and the scalars (the Higgs multiplet) are also  part of the approximately supersymmetric dynamics.
Probably the most efficient way to proceed is by employing    the proper generalization of the CCWZ \cite{Coleman:1969sm}  coset construction to the case of a non-linearly realized space-time symmetry (supersymmetry), which was laid down by Volkov and  Ogievetsky  \cite{ogievetsky, Volkov:1973vd} in the 70's, and for which a recent reappraisal can be found in Ref.~\cite{Delacretaz:2014oxa}.
A proper investigation of that construction is however beyond  the scope of our present discussion, and we leave it for future work. Here we will content ourselves by illustrating  qualitatively
which interaction structures  can possibly arise when ${\cal N}$ of the SM fermions are pseudo-Goldstini.
 That can be done by inspecting  \eq{goldststructuresN1}, which suggests which classes of terms to expect when   ${\cal N}>1$, while obviously the underlying $R$-symmetry will constrain the contraction of the $a$ indices of the Goldstini. As a matter of fact, we have also done a  more quantitative analysis, by  finding, through a Noether procedure, the most general Lagrangian and transformation laws at order $1/F^2$. In other words, we have
 explicitly constructed a Lagrangian and transformation laws that satisfy ${\cal N}$ non-linear supersymmetries up to terms of order higher than the first in the $1/F^2$ expansion.
In particular, we find interactions of the form 
\begin{eqnarray}
i\frac{\kappa^{AB}_{ab}}{F^2}F^{A\,\rho\mu}F_{\mu\nu}^B\bar \psi_a(\gamma_\rho\partial^\nu+\gamma^\nu\partial_\rho)\psi_b\,, 
\label{bigLag1}
\end{eqnarray}
whose action is invariant under $\delta\psi^a=\xi^a+\cdots$ and  $\delta F^A_{\mu\nu}=  i\frac{\kappa^{AB}_{ab}}{F^2}\partial_\mu(F^B_{\nu\tau}\bar{\psi}_a)\gamma^{\tau}\xi_b+\textrm{h.c.}+\cdots$ up to terms at least $O(1/F^4)$.\footnote{Of course, unless the supersymmetry is exact, these higher order terms will generate, via quantum effects, symmetry breaking interactions at all orders, including ${\cal O}_{4\psi}$ at dimension-6. Nevertheless, the leading contribution to ${\cal O}_{4\psi}$ comes from loops involving for instance a non-supersymmetric  operator of dimension 12 
and  will be suppressed by $\sim (g_*/4\pi)^4$ from closing  the loops to reduce the number of legs.} 
A relevant case arises when  $\kappa^{AB}_{ij}F^{A\,\rho\mu}F_{\mu\nu}^B=W^{a\,\rho\mu}B_{\mu\nu}\sigma^a_{ij}$. In that case  (a factor of) the underlying $R$-symmetry group is identified with the global subgroup of the SM gauge group factor $SU(2)_L$  and the field-strength couple to an isospin current, as shown in  \eq{ftt2} of appendix \ref{app:dim8}. 
That represents a first step towards the construction of an  action invariant under the non-linear transformation of ${\cal N}$ Goldstini, a task that we leave for future work. 

To conclude, we have sketched a situation where an approximate supersymmetry suppresses interactions of dimension-6 involving fermions and scalars/gauge bosons, in favor of  interactions of dimension-8 with more derivatives, which we summarize in Appendix \ref{app:dim8},  eqs.~(\ref{eqTvv}-\ref{ftt2},\ref{eqTH}).

\section{Applications to Weak-scale Effective Lagrangians}\label{sec:EW}

Based on the ideas of the previous section we shall here  present scenarios where all the SM degrees of freedom (fermions, gauge fields and the Higgs) can  take part in a novel  strong dynamics around  the TeV scale.
In those models where the gauge bosons are composite with 
 strong multipolar interactions,  the  symmetry of the strong sector will generally be
 \be
[{\cal G}]_{global}\times [U(1)^N]_{local}\,,
\ee
which will be deformed by turning on the gauge couplings into
\be
[{\cal H}_{1}]_{global}\times [{\cal H}_2]_{local}\,,
\ee
where ${\cal H}_1\times {\cal H}_2$ is a subgroup of ${\cal G}$ and ${\mathrm {dim}}\,[{\cal H}_2]=N$. Such a deformation, which slightly generalizes 
\eq{deform}, will allow us to implement the situation  where the Higgs is a PNGB.

For a number of reasons  we ended up labeling  \emph{Remedios} the  scenarios with composite vectors: the  vague esthetic analogy
with a  character of a famous novel, its r\^ole as a remedy to provide a physical interpretation of some LHC searches, and finally for the impossibility to provide an explicit UV realization.\footnote{
 \emph{Remedios the Beauty was not a creature of this world} - Gabriel Garcia Marquez.}

\subsection{Pure Remedios} \label{sec:rem}
The simplest scenario, though perhaps  not the most motivated,  is one  where only   the gauge fields  (or part of them) participate in a strong multipolar dynamics. In such a 
pure \emph{Remedios} only  the operators associated with the SM field-strengths $W^a_{\mu\nu}$,  $B_{\mu\nu}$ and $G^A_{\mu\nu}$ can   appear  enhanced by powers of the strong coupling $g_*$ in the effective Lagrangian. 
 Therefore the largest  effects   will be given by operators that are purely built from field-strengths and (covariant) derivatives. At the dimension-6 level  there are  just ${\cal O}_{3W}$, ${\cal O}_{3G}$  and    ${\cal O}_{2V}$ ($V=W,B,G$).
The coefficients of these operators are expected to be of order
\be
c_{3W},c_{3G}\sim g_*\ ,\ \ \ \ 
c_{2W},c_{2B},c_{2G}\sim 1\, .
\label{estim}
\ee
In the electroweak sector
the main effects  of these   dimension-6 operators
are anomalous triple gauge couplings (TGC), which affect  diboson production, 
and   modifications of the gauge boson propagators,  which affect  $\psi\bar \psi \to \psi\bar \psi$:
\bea
c_{3W}\sim g_* \quad &\Rightarrow& \quad  \delta {\cal A}(\bar\psi\psi\to V_TV_T)\sim g g_*\frac{E^2}{m_*^2}\ ,
\ \ \  \ \ (V_T=W_T,Z_T)\label{remffVV} \\
&&\ \ \ \delta{\cal A}(V_TV_T\to V_TV_T)\sim g g_*\frac{E^2}{m_*^2}, g_*^2\frac{E^4}{m_*^4}\, ,\label{remVVVV}\\
c_{2W}, c_{2B}\sim 1 \quad &\Rightarrow& \quad \delta{\cal A}(\psi\bar \psi\to V^*_T\to \psi\bar\psi)\sim g^2 \frac{E^2}{m^2_*} \,.
\label{remffff}
\eea
For $g_*\gg g$ the effects of $c_{2W,2B}$ are subleading  
 to those of $c_{3W}$. However the high precision 
with which  $e^+e^- \to V^*\to \psi\bar \psi$ was measured at LEP1/SLC  and LEP2 makes the former effects  
 more relevant at the present moment.
These effects are encapsulated by the   $W,Y$ parameters defined  in \cite{Barbieri:2004qk}:
\be
  W,Y\equiv c_{2W,2B} \frac{m_W^2}{m^2_*}\sim \frac{m_W^2}{m^2_*}\,.
  \ee  
Electron-positron  data imply   $W,Y\lsim { 10^{-3}}$  \cite{Barbieri:2004qk},
limiting this scenario  to  $m_*\gsim  3$ TeV. 
On the other hand, $c_{3W}$ contribute mainly to 
 TGC, which are presently  less well measured.
The effect  of $c_{3W}$
has  the same structure as $\lambda_\gamma$  defined in Ref.~\cite{Hagiwara:1986vm}:
\be
\lambda_\gamma\equiv c_{3W}\frac{m_W^2}{m^2_*}\sim g_*\frac{m_W^2}{m^2_*}\,,
\ee
whose  present experimental constraint  $\lambda_\gamma\lesssim  \textrm{few}\times10^{-2}$ \cite{Schael:2013ita,Falkowski:2014tna}   leads to  a weaker bound on these scenarios: $m_*\gtrsim { 1.5}\sqrt{g_*/(4\pi)}$ TeV. However the sensitivity will   improve with the advancement of the  LHC program \cite{other,Bian:2015zha}.

Finally, effects in ${\cal A}(V_TV_T \to V_TV_T)$ have just started to be under LHC scrutiny using different channels \cite{Aad:2014zda,CMS:2015jaa,CMS:2015ibc}. Nevertheless,
 a self-consistent analysis including effects from higher-dimensional operators  is still missing.
 In this respect, it is interesting to notice that our estimates in \eq{remVVVV}
 reveals a second contribution to ${\cal A}(V_TV_T \to V_TV_T)$ proportional to  $g^2_* (E/m_*)^4$. Besides coming from
two insertions of $c_{3W}$, a similar contribution is given by
  dimension-8 operators  of the schematic form  $(g_*^2/m_*^4)F_{\mu\nu}^4$ 
(see Appendix \ref{app:dim8} for details).
    These contributions cannot be neglected and, in fact,  dominate over one single insertion of $c_{3W}$
    when $ E^2/m_*^2\gtrsim g/g_*$. Phrased differently, as soon as the dimension-6 contribution is  bigger than the SM one, the whole cross-section becomes sensitive to  dimension-8 effects.\footnote{Surprisingly, the peculiar helicity structure of these scattering amplitudes  forbids interference between dimension-6 operators and the SM in high-energy processes  involving at least one transversely polarized  gauge boson~\cite{WWWW}. For amplitudes with only transverse polarizations, this implies that   dimension-6 and dimension-8 operators will give comparable corrections $\propto g^2g_*^2 E^4/m_*^4$ to the squared amplitude.} 
 
 Analogous considerations apply when gluons $G^A_\mu$  are involved in the strong dynamics.

\subsection{Remedios with a Composite Higgs}

Since the need for a  strong dynamics at the TeV is mainly motivated by the hierarchy problem,
it is worth considering  strong sectors in which  also the Higgs arises as  a  composite state. Furthermore, 
the little hierarchy problem forces us to assume  the Higgs field is  a PNGB parametrized by a coset  $\cal G/H$.  That assumption  directly
translates into the  request  that the strong couplings of $W^a_{\mu\nu}$ and $B_{\mu\nu}$  preserve $\cal G$. Absent that request  the Higgs mass
would be destabilised by the strong dynamics.  
We can see two broad options to achieve that, as we illustrate in the next two sections.

\subsubsection{Minimal Composite Higgs Model}\label{so5remedios}
Let us consider the simplest scenario where $H$ is a PNGB multiplet describing the coset $SO(5)/SO(4)$ \cite{Agashe:2004rs}. In the standard construction, $SU(2)_L\times U(1)_Y$ gauges a subgroup of $SO(5)$, thus explicitly breaking the symmetry that ensures the masslessness of the Goldstones. As long as the weak gauge group is weak, the breaking of the symmetry is small and consequently the Higgs mass is smaller than the fundamental scale $m_*$, a necessary condition for the theory to be not badly tuned. However, if,  in the limit where $g=g'=0$, the vector bosons interact strongly via multipolar terms, they should do so respecting $SO(5)$. In particular, that implies  $W^a_\mu$ must necessarily be a triplet under some additional $\widetilde{SU}(2)\subset\!\!\!\!\!\not\hspace{3mm} SO(5)$. 

Starting from the Minimal Composite Higgs Model (MCHM), the simplest option is then to consider a strong sector which, in the $g=g'=0$ limit,
has the symmetry group
\be 
{\cal G}=[SO(5)\times \widetilde{SU}(2)\times U(1)_X]_{global}\times [U(1)^4]_{local}\, , \label{so5su2}
\ee
the four local $U(1)$'s being associated with the three $W^a_\mu$ and $B_\mu$. Under the global group,
 $W_\mu^a$ transforms as a triplet $(\bf{1},\bf{3},{0})$, while $B_\mu$ is a total singlet. When $g$ and $g'$ are turned on, the group  \eq{so5su2} is {\it {deformed and reduced}}, following the inverse of the IW contraction of 
 Section~\ref{sec:dipolestrong}. In particular, given $SU(2)_L\times SU(2)_R\cong SO(4)\subset SO(5)$, $g $ gauges the diagonal $SU(2)$  inside $SU(2)_L\times\widetilde{SU}(2)$. On the other hand, $g'$ gauges, as in standard constructions, the linear combination $T_{3R}+X$.  The resulting effective Lagrangian has then the form 
 \be
 {\cal L}_{eff}=\frac{m_*^4}{g_*^2} L\left(U, \frac{\hat F^i_{\mu\nu}}{m_*^2},\frac{D_\mu}{m_*}\right)\,,
 \label{Lso5su2}
 \ee
 where $U=e^{i\Pi}$ is the Goldstone matrix that contains the Higgs and  ${\hat F_{\mu\nu}}$ and  ${D_\mu}$ are given by eqs.~(\ref{coder},\ref{Fmunu}), with $\epsilon =g/g_*$ and $\epsilon'=g'/g_*$ for respectively $SU(2)_L$ and $U(1)_Y$.  The effective Lagrangian will also be written as an expansion in loops, controlled by $(g_*/4\pi)^2$, and in powers of the symmetry breaking parameters $\epsilon$ and $\epsilon'$, as previously illustrated  in \eq{loopexp} and \eq{LNL}.
 The Lagrangian in \eq{Lso5su2} respects the symmetry   \eq{so5su2}, apart from the terms associated with $g$ and $g'$.   According to the discussion in Section~\ref{partcomp}, our Lagrangian  can also be formally thought  as arising in  a scenario  where  weakly coupled vectors are mixed to the strong sector via the terms
 \be
 \epsilon_W W_{\mu\nu}^a {\cal O}^{a\, \mu\nu}+\epsilon_B B_{\mu\nu} {\cal O}^{\mu\nu}\,,
 \ee
and then the  limit $\epsilon_{W,B}\to 1$ is taken.
 
 It is now straightforward to estimate the size of the different contributions to the effective Lagrangian: only operators that are invariant under ${\cal G}$ can  arise from the strong dynamics. 
  In particular, at dimension-6 the only such 
 operators are ${\cal O}_H$ and  ${\cal O}_{3W}$ for which we thus expect
  \be
 c_H\sim g^2_*\ , \ \ \ \  c_{3W}\sim g_*\,.
 \ee
On the other hand   ${\cal O}_W$ and  ${\cal O}_{HW}$ are not invariant under $\widetilde{SU}(2)$, while 
   ${\cal O}_{B}$ and ${\cal O}_{HB}$  are forbidden by  $SO(4)$
(since  $H^\dagger \lra D_\mu H$ and  $D_{[\mu} H^\dagger  D_{\nu]} H$ belong to $\bf (1,3)$ of $SU(2)_L\times SU(2)_R$). 
   The coefficients of those other operators  
will therefore be  suppressed by powers of the weak couplings, like in ordinary composite Higgs scenarios \cite{Giudice:2007fh}.  It perhaps does not make much sense to assume  here MC (see Appendix~\ref{minimalcoupling}), as, unlike for the case of  the standard composite Higgs scenario, we do not possess here  
any explicit, if partial,  UV completion, and even less  one that satisfies MC.
Finally, the coefficients $c_{2W}$ and $c_{2B}$ are of $O(1)$,  as  in  \eq{estim},
contrary to the SILH where they are suppressed by $O(g^2/g_*^2)$.
A summary of these results is given in  table~\ref{operatorsdim6coeff}.

On the other hand, at the dimension-8 level,  in addition to the $F_{\mu\nu}^4$ terms discussed above,
 further  operators  with  $O(g_*^2)$ coefficient will arise. In particular, we shall have operators involving two $H$ and two
  field-strengths,  e.g., $\epsilon^{abc}D^\mu H^\dagger\sigma^a D_\nu H\, W^b_{\mu\rho}  W^{\nu\rho\, c}$ (see Appendix \ref{app:dim8}).  These operators, however, only contribute  to the scattering of four bosons,  through $O(g_*^2)$ corrections to  $VVVV$,  $VVhh$ and $V^3h$ vertices.

 \begin{table}[t]
\renewcommand{\arraystretch}{1.2}
{
\begin{center}
\begin{tabular}{|c|c|c|c|c|c|c|c|c|c|}
\hline 
Model &${\cal O}_{2V}$ & ${\cal O}_{3V}$ & ${\cal O}_{HW}$ & ${\cal O}_{HB}$ & ${\cal O}_{V}$ & ${\cal O}_{VV}$ & ${\cal O}_{H}$ & ${\cal O}_{y_\psi}$      \\
\hline
%
{\footnotesize Remedios (sect. \ref{sec:rem}) } & $1$ & $g_*$ &  &  &  & & &   \\
{\footnotesize  Remedios+MCHM  (sect. \ref{so5remedios}) }& $1$ & $g_*$ &$g$       & $g'$ &
$g_V$ & $g_V^2$ & $g_*^2$  & $y_\psi g_*^2$  \\
{\footnotesize  Remedios+$ISO(4)$  (sect. \ref{iso}) } & $1$ & $g_*$ & $g_*$  & $g'$   &  $g_V$   & $g_V^2$   & $ \lambda_h$&   $y_\psi\lambda_h$ \\
\hline
\end{tabular}
\end{center}
}
\caption{\emph{Estimated coefficients of the  dimension-6 operators   for the different scenarios considered in the main text,
neglecting loop effects.
The subscript $V$ can denote  $W,B,G$ according to  the basis defined in table~\ref{operatorsdim6}. 
}}
\label{operatorsdim6coeff}
\end{table}%

\subsubsection{Remedios with $ISO(4)$}\label{iso}

If we are willing to possibly give up  UV-completions within quantum field theory (as explained, this might even have to be the case for \emph{Remedios} models), we can consider  scenarios in which  the Higgs is a PNGB issued from the spontaneous breakdown of a non-compact group.  The simplest such option is given by $SO(4,1)/SO(4)$, see for instance the discussion in \cite{Contino:2013gna}\footnote{We can write a perfectly consistent effective Lagrangian based  on
 $SO(4,1)/SO(4)$, given the unitarity of the residual symmetry $SO(4)$. However
a  unitary CFT cannot respect a non-compact symmetry such as $SO(4,1)$: within ordinary QFT, there  cannot exist a UV-completion with linearly realized $SO(4,1)$.}.  However, its   combination with \emph{Remedios} does not introduce novelties with respect to $SO(5)/SO(4)$, as we still need  to add an additional ${\widetilde {SU}}(2)$ to include the 
$W^a_\mu$ in the strong dynamics.
A more interesting example is to consider   $ISO(4)$, the isometric group of  the 4D Euclidean space,
consisting of the semidirect product of 4D rotations and translations: $SO(4)\rtimes T_4$. The Higgs field can here be identified with  the coset $ISO(4)/SO(4)$  corresponding to the transformation
\bea
&&H\rightarrow H+c\, ,\nonumber\\
&&H\rightarrow RH\, ,
\eea
where $c$ parametrizes $T_4$ translations while $R$ is an $SO(4)$ rotation of the 4-component Higgs.\footnote{The same reasoning of the previous footnote applies: there cannot exist any ordinary unitary QFT with linearly realized $ISO(4)$ acting as UV completion.}

Now, with respect to $SO(5)$, the important novelty is  that $ISO(4)=SO(4)\rtimes T_4$ possesses irreducible representations of dimension 3. These are the $(\bf{3},\bf{1})$ and $(\bf{1},\bf{3})$ of the $SU(2)_L\times SU(2)_R$.
We can thus fit the triplet of electroweak bosons $W^a_\mu$ in the $(\bf{3},\bf{1})$ while respecting the full $ISO(4)$.

Concerning instead the hypercharge field $B_{\mu}$, the simplest option is to take it as a singlet, as in the previous example. That choice obviously  preserves $ISO(4)$, in particular the Higgs would remain a Goldstone. Alternatively, $B_{\mu}$ could be fit into an incomplete $(\bf{1},\bf{3})$ of $SO(4)$. That  would break   $SO(4)$ down to $ SU(2)_L\times U(1)_R$, but would leave $T_4$ untouched. In practice the coset would become $[SU(2)\times U(1)\rtimes T_4]/SU(2)\times U(1)$: the Higgs would remain a Goldstone, but  the custodial symmetry would be maximally broken. In what follows we shall focus on the first option, for which the strong dynamics, in the pure Remedios limit,  has symmetry group
\be 
{\cal G}=[ISO(4)]_{global}\rtimes [U(1)^4]_{local}\, , \label{iso4}
\ee
with the gauge bosons of $U(1)^4$ transforming like a $(\bf{3},\bf{1})\oplus (\bf{1},\bf{1})$ under $SO(4)$. Moreover, $ISO(4)$ is assumed to be spontaneously broken to $SO(4)$.

Before discussing the effective Lagrangian and its implications, it may be worth recalling that $ISO(4)$ is a IW contraction of both $SO(5)$ and $SO(4,1)$. Geometrically this is quite evident: the isometries of the plane ($ISO(N)$) coincide with the local  limit of the isometries of the sphere ($SO(N)$), or of any hyperboloid ($SO(N-m,m)$). As a consequence of that, the effective Lagrangian for $ISO(4)/SO(4)$ can be viewed as a particular limit of that for $SO(5)/SO(4)$. In the CCWZ language~\cite{Coleman:1969sm} (and in   the notation of ref.\cite{Weinberg:1996kr}), that amounts to replacing everywhere $d_\mu$ and $E_\mu$ according to
\be\label{CCWZISO}
d_\mu^\alpha \hat T_\alpha +E_\mu^A T_A=-ie^{-i\pi^\alpha\hat T_\alpha}\partial_\mu e^{i\pi_\alpha\hat T_\alpha}\longrightarrow \frac{-i}{\epsilon} e^{-i\epsilon\pi^\alpha\hat T_\alpha}\partial_\mu e^{i\epsilon\pi_\alpha\hat T_\alpha}\equiv d_\mu^\alpha(\epsilon) \hat T_\alpha +E_\mu^A(\epsilon) T_A\,,
\ee
and then letting $\epsilon \to 0$, so that $d_\mu^\alpha\to \partial_\mu \pi^\alpha$ and $E_\mu^A\to 0$. This specific result for  $d_\mu$ and $E_\mu$ simply corresponds to the metric flatness of the $ISO(4)/SO(4)$ coset. One important consequence is that in the $\cal G$ symmetric limit, that is for $g=g'=0$, the effective Lagrangian only depends on derivatives of the Higgs field: $L\equiv L(\partial H)$.

Invariance under ${\cal G}$ strongly restricts the   interactions generated by the strong dynamics.
Among the dimension-6 operators of table~\ref{operatorsdim6}, only ${\cal O}_{3W},\, {\cal O}_{HW},\,
{\cal O}_{2W},\,{\cal O}_{2B}$ have   coefficients enhanced according to the pattern 
\be
 c_{3W},c_{HW}\sim g_*
\ ,\ \ \ 
c_{2W},c_{2B}\sim 1 \, .
\label{estim2}
\ee
 In particular,  the shift symmetry $T_4$ implies the constraint $c_H=0$. Indeed $c_H$ controls the curvature of the coset manifold: it is positive for $SO(5)/SO(4)$, negative for $SO(4,1)/SO(4)$ and vanishing for $ISO(4)/SO(4)$. Similarly,
  $T_4$ forbids direct strong-sector contributions to  $c_W$, $c_B$, $c_T$, which nicely helps control corrections to  electroweak observables, while  $SO(4)$  forces $c_{HB}=0$. The operator  
$|\Box H|^2$, as it
respects ${\cal G}$, can appear with $O(1)$ coefficient. However, by use of the equations of motion  it can be easily seen to only affect ${\cal G}$-invariant interactions of dimension greater than 8. Its fate is however different when the small deformations of ${\cal G}$ are taken into account, as we shall discuss 
later. Finally, as in the previous examples, at dimension-8 level we encounter new operators with  $O(g_*^2)$ coefficients, which contribute to  $VV\to VV$  at $O(g_*^2E^4/m_*^4)$, see Appendix \ref{app:dim8}, \eq{dim8hhscatt} in particular.

\vskip0.6truecm
\noindent
{\bf $\cal G$-breaking Effects and Phenomenology}
\vskip0.2truecm
\noindent
In order to obtain a low energy  description coinciding with the SM, we must  necessarily  deform $\cal G$. In particular $ISO(4)$ must be explicitly broken. There are in principle multiple model building options for the structure of that breaking, at each order in the derivative and field expansion in the effective Lagrangian. Moreover an important structural factor concerns the possible involvement of the fermions in the strong dynamics. Here we shall make the conservative and consistent  assumption that all the sources of ${\cal G}$-breaking  are associated with  the $SU(2)_L\times U(1)_Y$ gauge couplings, the  Yukawa couplings and the Higgs potential. Moreover we shall take  the fermions to be elementary, that is we shall assume their interactions 
 with the strong sector are all associated with the (weak) couplings of the SM. On the other hand, we will remain agnostic onto the source of these couplings. In particular we shall not rely on partial compositeness, which provides a direct way to estimate the coefficients of symmetry breaking operators of arbitrary dimension.
Nonetheless,  we find it more than reasonable to assume that, whatever the underlying theory, each and every SM interactions is accompanied by a tower of higher derivative interactions involving precisely the same fields and suppressed  just by the suitable powers of $1/m_*$  with respect to the lower derivative terms. As it will turn out, this tower of higher derivative operators is  the leading source of ${\cal G}$-breaking higher dimension operators. The other source is  given by quantum corrections saturated at the scale $m_*$. These effects are unescapable unless some fine tuning is allowed. They thus provide 
 a lower bound on the size of the operator coefficients.

According to the above discussion, aside the gauge couplings $g\equiv \epsilon g_*$ and $g'\equiv \epsilon' g_*$, we shall consider as leading
 $\cal G$-breaking effects the top Yukawa and the Higgs potential
\be
{\cal L}_{break}= \epsilon_t g_* \left [\bar Q_LHt_R+\dots\right ]+\epsilon_2 m_*^2 \left [|H|^2+\dots\right ] -\epsilon_4 \frac{g_*^2}{2}\left [|H|^4+\dots\right ]\,,
\label{break}
\ee 
where for each of the three expression in square brackets, the dots represent higher derivative operators with the same field content and purely suppressed by powers of $1/m_*$.
Here we should identify  $\epsilon_t g_*\equiv y_t$, $\epsilon_2 m_*^2\equiv m_H^2$ and $\epsilon_4 g_*^2\equiv \lambda_h$ with the corresponding SM coupling renormalized at the scale $m_*$. Notice that in the interesting case where $m_*$ is not much larger than a few TeV, renormalization effects
are typically not very important. A relevant exception is given by the Higgs quartic, which receives a sizeable IR contribution from top quark loops
\be
\Delta\lambda_h = \frac{3y_t^4}{4\pi^2}\ln \frac{m_*}{m_t}\,. \label{deltalambda}
\ee
As is well known from the case of the MSSM, for $m_*$ of order a few TeV, and given the observed value of the Higgs mass, the IR and UV contributions are comparable: $\Delta \lambda_h\sim \lambda_h$. We shall thus make this rough identification throughout our estimates.

Aside calculable IR effects, the breaking of $ ISO(4)$ in eq.~(\ref{break}) will propagate through quantum corrections at the cut-off scale $m_*$, of which we can only offer an order of magnitude estimate, and whose detailed value  depends on the full theory.  First and foremost we have the top sector correction to the Higgs 
mass parameter, expected to scale like 
\be
\Delta m_H^2 \sim \frac{3y_t^2}{4\pi^2} m_*^2\,.
\ee
That implies the usual estimate of the tuning
\be
\frac{(125 \GeV)^2}{\Delta m_H^2}\sim \left (\frac{400 \GeV}{m_*}\right )^2\, ,
\ee
or equivalently, given $\lambda_h\sim \Delta \lambda_h$ and eq.~(\ref{deltalambda}),
\be 
\frac{\lambda_h v^2 }{\Delta m_H^2} \sim \frac{m_t^2}{m_*^2} \ln \frac{m_*}{m_t}\equiv \frac{v^2}{{\tilde f}^2}  \ln \frac{m_*}{m_t}\, .
\label{naturalvev}
\ee
In our scenario, $\tilde f \equiv m_*/y_t$ thus represents  the natural scale for $v$, the Higgs   vacuum expectation value (VEV). \footnote{It is interesting to compare this results
with ordinary composite Higgs models based on a compact coset, like $SO(5)/SO(4)$. In that case the natural scale for non linearities in the Higgs field
and thus for the Higgs VEV is $f\equiv  m_*/g_*$. 
Instead, in a flat non-compact coset like $ISO(4)/SO(4)$ the quantity $m_*/g_*$ does not represent the natural scale for the   VEV: in the unbroken theory all points in the infinite $H$-plane are equivalent. Only when $\epsilon_i$ are turned on, does there  arise a natural scale for the  VEV, and given by $\tilde f$.}

We can now estimate the size of the coefficients of dimension-6 operators that arise when ${\cal G}$ is deformed by the SM couplings:
\begin{itemize}
\item The first class of operators   involves four derivatives and two powers of the Higgs field, 
schematically of the form $H^\dagger \partial^4 H/m_*^2$. When the partial derivatives $\partial_\mu$ are deformed into convariant derivatives $D_\mu$, the resulting structures are classified according to the number of  commutators, $[D_\mu,D_\nu]=-igW^a_{\mu\nu}\sigma^a/2-ig^\prime B^a_{\mu\nu}$, they involve. These resulting operators will thus involve 0, 1 or 2 field-strengths, each weighted by the corresponding gauge coupling.
 There is just one operator not involving any field-strength, and that is $|\Box H|^2/m_*^2$,  which respects ${\cal G}$ and which we already encountered in the discussion of the ${\cal G}$-symmetric limit. In the presence of the SM couplings (eq.~(\ref{break})) its consequences are however different. By applying a field redefinition this operator can be rewritten in terms of the operators in table~\ref{operatorsdim6}. In particular, one finds
 $c_6\sim \lambda_h^2$, $c_{4\psi}\sim y_\psi^2$ and more importantly a  contribution to $c_{y_\psi}$
 that gives a universal correction to the Higgs couplings to fermions:
\begin{equation}\label{eqiso1}
c_{y_\psi}\sim y_\psi\lambda_h \quad \Rightarrow \quad  \delta g_{h\psi\psi}\sim    \frac{m_h^2}{m_*^2},
\end{equation}
where $\Delta{\cal L}^h_{\psi\psi}=(h/v)(\delta g_{h\psi\psi}m_\psi\bar\psi\psi+\textrm{h.c.})$. 
Concerning operators involving field-strengths, by a straighforward analysis one can prove they give rise to independent contributions to 
${\cal O}_{HB},\,{\cal O}_W, \,{\cal O}_B,\, {\cal O}_{BB}$. In particular that implies  $c_{HB}\sim g^\prime$ and, more interestingly,
\begin{align}
c_{B}\sim     g' \, , \,\,\,\,c_{W}\sim  g \quad &\Rightarrow \quad   \delta \widehat S\sim   \frac{m_W^2}{m_*^2}\ ,
\label{estim3}
\end{align}
and
\begin{equation}\label{estim4}
c_{BB}\sim  g^{\prime\, 2}\quad \Rightarrow \quad   \delta g_{h\gamma\gamma}\sim   \frac{e^2 v^2}{m_*^2}\ ,
\end{equation}
where  $\widehat S$ (and for later $\widehat T$) are defined in \cite{Barbieri:2004qk} and are proportional to the Peskin-Takeuchi parameters\cite{Peskin:1990zt}, while modifications to the Higgs coupling to photons are normalized as $\Delta{\cal L}^h_{\gamma\gamma}=(h/v)\delta g_{h\gamma \gamma} F_{\mu\nu}F^{\mu\nu}$.
Notice that  both  $c_B$ and $c_{HB}$  break
the custodial $SO(4)$, while $c_B$ also breaks $T_4$. Nevertheless, one  insertion of $g'$  saturates the necessary selection rule. 
As a result the relative size of $c_B\sim c_{HB}\sim (g^\prime/g)  c_W$ from \eq{estim3} is the same as in the SILH (without MC), while  $c_{HW}$ is enhanced by $\sim g_*/g$ (see eq.~(\ref{estim2})).

\item A second class of effects results from the dressing of ${\cal G}$-breaking interactions with  powers of $\partial_\mu/m_*$, as captured by the dots in \eq{break}. 
In particular, ${\cal O}_H$ can be viewed as just a 2-derivative iteration of the Higgs quartic term, protected by the same selection rules, and thus controlled by the same small parameter  $\epsilon_4 g_*^2\equiv \lambda_h $:
\begin{equation}
c_{H}\sim \lambda_h\quad \Rightarrow \quad   \delta g_{hVV}\sim  \frac{m_h^2}{m_*^2}\ , \\
\end{equation}
where $\Delta{\cal L}^h_{VV}=(h/v) \delta g_{hVV}m_W^2 \left(W^{+\mu}W^-_\mu+\frac{Z^{\mu}Z_\mu}{2\cos\theta_W}\right)$ parametrizes deviations of the Higgs couplings to vectors. Similarly, higher derivative dressing of the Yukawa 
couplings, $y_\psi \bar \psi_L \psi_R  \Box H$,
 can be shown to  lead, upon use of the $H$ equation of motion,
to $c_{y_\psi}\sim y_\psi\lambda_h$. Although these other effects are comparable to \eq{eqiso1},  they are distinguished, as they are in general  not universal across fermion species.

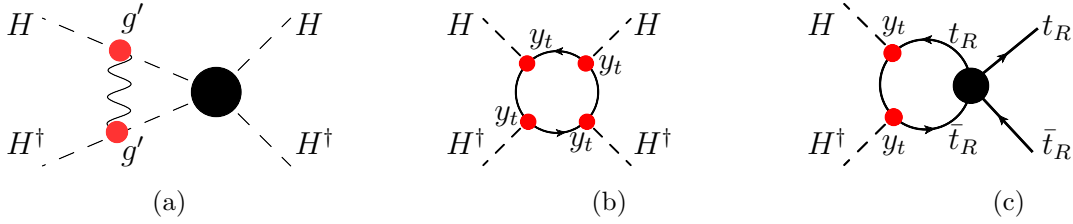
\begin{figure}[t]
\begin{center}
\hspace{1cm}\subfigure[]{\begin{picture}(100,80)
\put(0,0){\includegraphics[height=2cm]{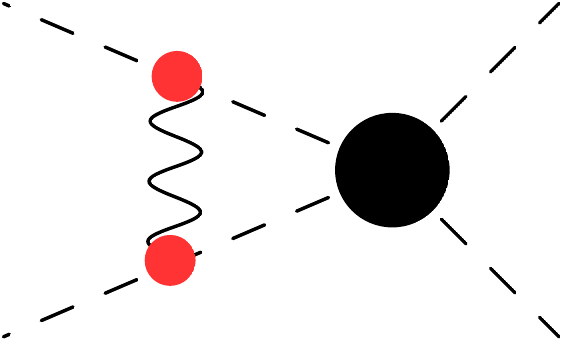} }
\put(-13,50){$H$}
\put(-13,5){$H^\dagger$}
\put(30,53){$g^\prime$}
\put(30,3){$g^\prime$}
\put(95,50){$
H$}
\put(95,5){$
H^\dagger$}
\end{picture}\label{figd}}
\hspace{2cm}
\subfigure[]{\begin{picture}(100,80)
\put(0,0){\includegraphics[height=2cm]{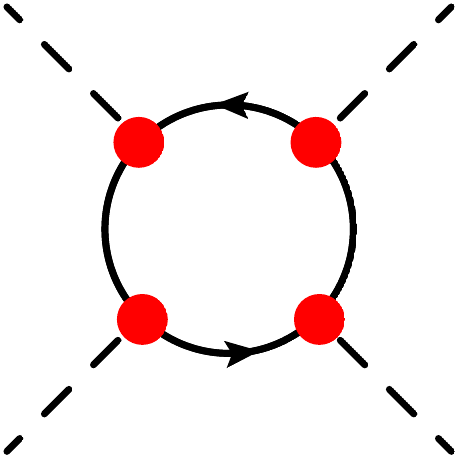} }
\put(-13,50){$H$}
\put(-13,5){$H^\dagger$}
\put(18,47){$y_t$}
\put(44,37){$y_t$}
\put(33,8){$y_t$}
\put(5,18){$y_t$}
\put(57,50){$H$}
\put(57,5){$H^\dagger$}
\end{picture}\label{fige}}
\hspace{.5cm}
\subfigure[]{\begin{picture}(155,80)(-13,0)
\put(0,0){\includegraphics[height=2.2cm]{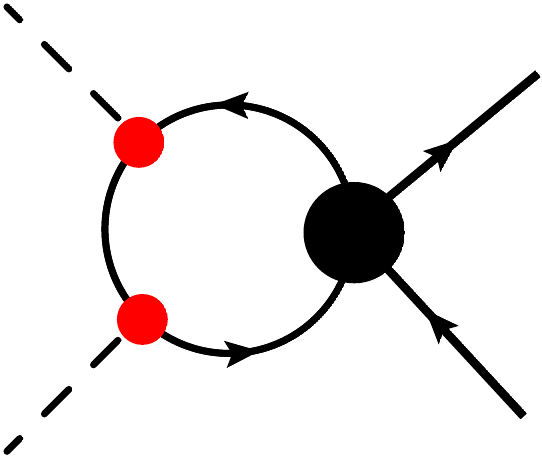} }
\put(-13,50){$H$}
\put(-13,5){$H^\dagger$}
\put(15,52){$y_t$}
\put(15,7){$y_t$}
\put(40,47){$t_R$}
\put(40,7){$\bar t_R$}
\put(75,50){$t_R$}
\put(75,5){$\bar t_R$}
\end{picture}\label{fig:FD2}}
\end{center}
\caption{\emph{{Diagrams a and b: leading contributions to the operator ${\cal O}_{T}$. Diagram c: leading effects in the case of  $\,t_R$ compositeness (for $t_L$ compositeness with the obvious replacement $R\to L$).
The black  blob denotes   $O(g_*^2)$  vertices from the ${ \cal G}\equiv[ISO(4)]_{global}\rtimes [U(1)^4]_{local}$ preserving strong dynamics at 
loop momentum virtuality $\sim m_*$. 
 The small red blobs denote instead vertices associated with the weak deformation and breakdown of $\cal G$  by the SM couplings~($g,\,g',\,y_t,\,\lambda_h$).}}}\label{fig:FD}
\end{figure}

\item A third class of  ${\cal G}$-breaking operators is generated by loop-effects.  These are typically sub-leading, except for $c_T$, which might not receive any tree-level contribution
as it corresponds to a violation of two units of custodial isospin, $\Delta I_c=2$, that is not present in \eq{break}.
At the one-loop level, however,  this  custodial   breaking  can be achieved either by two  insertions of $g'$ ($g'$ can be assigned to a $(\bf{1},\bf{3})$ of $SO(4)$) or by four  insertions of $y_t$ ($y_t$ can be assigned to a $(\bf{1},\bf{2})$).
 A contribution in the first class is given by diagram Fig.~\ref{figd}: 
\begin{equation}
c_{T}\sim \left(\frac{g_*}{4\pi}\right)^2\times g^{\prime 2}\quad \Rightarrow \quad   
\delta\widehat T\sim  \left (\frac{g_*}{4\pi}\right)^2\times  \tan^2\theta_W\frac{m_W^2}{m_*^2}
\, ,
\label{extrat}
\end{equation}
while a  contribution in the second class  is instead given by diagram Fig.~\ref{fige}, and simply 
corresponds to  a  deformation  at virtuality of order $m_*$ of the 
SM  top loop contribution to the $\rho$-parameter:
\begin{equation}\label{eqisolast}
c_{T}\sim  \frac{y_t^4}{16\pi^2}\quad \Rightarrow \quad   
\delta\widehat T\sim  \left (\frac{y_t}{4\pi}\right)^2\times \frac{m_t^2}{m_*^2}
\, .
\end{equation}
This second contribution, even if not enhanced by the strong dynamics
could be as important as \eq{extrat}, especially for $g_*$ smaller than $4\pi$.

\end{itemize}

In summary, the main phenomenological effects in $ISO(4)$ models are captured by eqs. (\ref{estim2},\ref{eqiso1}-\ref{eqisolast}). We find that the  Higgs coupling to photons, $\delta g_{h\gamma\gamma}$ from \eq{estim4}, is  a factor $\sim (4\pi/y_t)^2$ larger than in the SILH~\cite{Giudice:2007fh}, and  gives a rather strong constraint in this type of models:  $m_*\gsim 2 $ TeV from a 2$\permil$ measurement of $\delta g_{h\gamma \gamma}$ \cite{cmsandatlas}. A comparable constraints is also given by the $\widehat S$ parameter from \eq{estim3}, while $\widehat T$ seems slightly less important because of the $\tan^2\theta_W$ and loop suppressions.
On the other hand, corrections to the  Higgs coupling to vectors and fermions, $\delta g_{hVV}$ and $\delta g_{h\psi\psi}$, are    suppressed compared to the case of a generic composite Higgs scenario, and provide, at the moment, weaker constraints.

Nevertheless, 
while contributions to the observables of eqs. (\ref{eqiso1}-\ref{eqisolast})  are ubiquitous in new physics scenarios,  the $g_*$-enhanced effects \eq{estim2} are unique to our model. 
Apart from  $c_{3W}\sim g_*$, which is present  in all Remedios scenarios, we  now also have  $c_{HW}\sim g_*$, which mainly affects TGC  and  the $h Z\gamma$ vertex:
\bea
\delta g_1^Z &=& \frac{\delta \kappa_\gamma} {\cos^2\theta_W}  =\frac{\delta g_{hZ\gamma}} {\sin\theta_W \cos\theta_W}
=
 -\frac{m^2_Z}{m_*^2}  \frac{c_{HW}}{g}\sim  \frac{m^2_Z}{m_*^2}  \frac{g_*}{g}\ ,\label{corr}\\
  \lambda_\gamma &=& \frac{m^2_W}{m_*^2} \frac{c_{3W}}{g}\sim 
  \frac{m^2_W}{m_*^2} \frac{g_*}{g}\, ,
\eea
where  $\delta g_1^Z, \delta \kappa_\gamma, \lambda_\gamma$ correspond to anomalous TGC, normalized according to Ref.~\cite{Hagiwara:1986vm},\footnote{We recall that,  at the dimension-6 level,  the other parameters are fixed: $\lambda_Z =    \lambda_\gamma$ and 
$\delta \kappa_Z= \delta g_1^Z - \tan^2 \theta_W  \delta \kappa_\gamma$.} while
 $\delta g_{h Z\gamma}$ describes the anomalous  $HZ\gamma$ vertex  according to 
$\Delta{\cal L}_{hZ\gamma}=\delta g_{h\gamma Z}(h/v) F_{\mu\nu}Z^{\mu\nu}$. Notice the interesting correlation between  $\delta g_1^Z$, $\delta \kappa_\gamma$ and $\delta g_{hZ\gamma}$,
which could single out these scenarios if  deviations from the SM predictions  were to appear in the future.
Taking into account the present constraint from $h\to\gamma\gamma$ and EWPT, the allowed size of these other effects could still be as large as a few per-cent which is within the reach of    the  ongoing  LHC run,  or of  future electron-positron colliders.
It is important  to point out  these new  contributions to   $hZ\gamma$ are  
potentially  even larger than  the SM  one, as that  arises at one-loop. 
Perhaps more importantly, the relative size of the new physics effects in $hZ\gamma$ and in $h\gamma\gamma$ is \begin{equation}
\frac{\delta g_{h\gamma Z}}{\delta g_{h\gamma \gamma}}\sim \frac{g_*}{g}\,.
\end{equation}
Thus, deviations   larger than $O(1)$ in  $BR(h\to Z\gamma)$ are possible  at the moment.

Like  in all models with composite Higgs, it is also worth considering the situation where one, or both, of the chiralities of the top quark is part of the strong dynamics.
Consider first the case where $t_R$ is composite. One class of important effects is  again given by $\partial_\mu/m_*$ dressing of SM operators. The leading effect at the dimension-6 level is given by  $\bar t_R \!\not\!\! D^3 t_R/m_*^2$, which upon use of the equations of motion   leads to $c_{y_t}\sim y_t^3$  and to $c^t_{L}\sim c^{(3)t}_{L}\sim y_t^2$ corresponding to a specific linear combination of ${\cal O}^{t}_L$ and ${\cal O}^{(3) t}_L$. The first effect  corresponds to an  $(m_t/m_h)^2$ enhancement  w.r.t. eq.~(\ref{eqiso1}).
The second  one gives a  relative correction to the $Zt_L\bar t_L$ vertex of order $m_t^2/m_*^2$, leaving the $Zb_L\bar b_L$ vertex unaffected.
 The other class of contributions arise from loops, and the leading one is   shown in Fig.~\ref{fig:FD2}. This
gives rise, among other subleading effects,  to 
$c_{R}^t\sim  (g_*/4\pi)^2 y_t^2$, implying a relative correction  to the $Zt_R\bar t_R$ vertex
of order $(g_*/4\pi)^2 (m_t/m_*)^2$.
Analogous effects are generated in the case  of a composite $ t_L$,
(Fig.~\ref{fig:FD2}, with $t_R \to t_L$). Now the novelty is that also the $Zb_L\bar b_L$ vertex  is modified by a relative amount $(g_*/4\pi)^2 (m_t/m_*)^2$, implying the rather strong constraint $m_* \gsim 5\,(g_*/4\pi) \TeV$.

\subsection{Composite Fermions}

 In this section we shall discuss the implications of fermion compositeness, in its different incarnations of partial compositeness, described in Section~\ref{sec:partComp}, and approximate supersymmetry, described in Section~\ref{sec:stronglfermions}.

\subsubsection{ Partially Composite Fermions}\label{sec:compoferm1}
\label{seccf}

In Section~\ref{sec:partComp} we discussed how to incorporate composite fermions in a strongly coupled sector by
taking the limit in which  the effective mixing parameter $\epsilon_\psi$ in  \eq{mixingJO}
becomes $O(1)$ at the IR scale  where the CFT develops a mass-gap. 
If the SM fermions are composite,  
the main BSM effects  are associated  with 4-fermion contact-interactions, described by ${\cal O}_{4\psi}$. 
In order to avoid severe constraints from flavor-violating processes,  these interactions should preserve an approximate family symmetry. If  that is the case,  one of the  most important process to unravel composite fermions  is
  $\psi\psi\to \psi\psi$ at high energy, as its amplitude receives an energy-growing contribution from 
${\cal O}_{4\psi}$:
\begin{equation}
\delta{\cal A}(\psi\psi\to\psi\psi)\simeq \epsilon^4_\psi g_*^2\frac{E^2}{m_*^2}\,.
\label{4f}
\end{equation}
At the LHC these effects are  being searched in   the high-energy angular distributions of dijets, for quark compositeness, and in the spectrum of Drell-Yan processes when also the leptons are composite. In the former case, the leading contribution is from  $qq$ initial states, as these  have larger PDFs  than  $q\bar q$. In both cases, the data from LHC run 1 imply $m_*\gtrsim  (g_*\epsilon^2_\psi/4\pi) \times 60\TeV$ (see table~5 of Ref.~\cite{Domenech:2012ai}, and also ~\cite{RSW,Alves:2014cda,Pomarol:2013zra,deBlas:2013qqa}).
The lower bound on $m_*$ seems quite stringent, but it is quickly  reduced if fermions are not fully composite,  that is for  $\epsilon_\psi<1$.
Indeed,  provided the Higgs field  is fully composite,  stronger constraints can arise, for sufficiently small  $\epsilon_\psi<1$,
 by considering processes involving bosons along with fermions. 
In particular, the new physics contribution to  $\psi\bar \psi\to V_LV_L/hV_L$,
scales like $\sim \epsilon_\psi^{2}g_*^2 E^2/m_*^2$,  which,  compared to eq.~(\ref{4f}), features fewer powers of 
$\epsilon_\psi$. 

Here we want to elaborate more on such scenarios, focussing for simplicity on the case where the  Higgs is composite and the fermions are partially composite, while the transverse polarizations of gauge bosons are elementary (that is, we here do without the  Remedios construction).
To analyze   which are the most relevant observables to test these scenarios,
we adopt  a   power-counting  following  \eq{mixingJO}.
One is easily convinced that, as long as $\epsilon_\psi> g/g_*$, the  operators with the largest coefficients are those with the largest number of Higgs or fermion fields (as opposed to field-strengths and derivatives),
as they  directly probe the strong dynamics. 
We can divide these operators into 3  classes (see table~\ref{operatorsdim6} for notation):\footnote{We focus on operators invariant under the full family symmetry $SU(3)^5$. This is the set of operators that will definitely be present whatever our assumptions  about the origin of flavor, in particular, whatever our assumption about the underlying flavor symmetry.
Moreover, this implies that operators that change the chirality of the fermions ($\propto \bar \psi_L\psi_R$)
are proportional to the Yukawa couplings and can then be neglected, with the exception
of  ${\cal O}_{y_\psi}$ whose impact on Higgs physics can still be important.} 
a)  ${\cal O}_{4\psi}$, whose main impact was discussed above, 
b) operators that only affect   Higgs physics, ${\cal O}_H$, ${\cal O}_6$, ${\cal O}_{y_\psi}$, and c) the 8 independent operators
\be
{\cal O}_L^q\, ,\ \
{\cal O}_L^{(3)\, q}\, ,\ \
{\cal O}_R^u\, ,\ \
{\cal O}_R^d\, ,\ \
{\cal O}_L^l\, ,\ \
{\cal O}_L^{(3)\, l}\, ,\ \
{\cal O}_R^e\, ,\ \
{\cal O}_T,
\label{opesix}
\ee
which, in the vacuum $H=(0,v/\sqrt{2})$,
give modifications to  electroweak observables and are already  constrained by pre-LHC measurements. 
Parametrically, all  coefficients of the operators of \eq{opesix}  are of order  $\sim\epsilon_\psi^2 g^2_*$, with the exception of ${\cal O}_T$
whose coefficient depends on the details of   custodial symmetry breaking 
(here we are not assuming any particular composite Higgs model).
The bounds on these coefficients  come  mainly from precision  $Z$-physics at LEP. 
In particular,  the  seven partial widths $\Gamma(Z\to\bar\psi\psi)$, with $\psi=e_{L,R},\nu_L,u_{L,R},d_{L,R}$, roughly place a constraint $m_*\gtrsim  (g_*\epsilon_\psi/4\pi) \times 40\TeV$  on seven linear combinations of the coefficients of the  8 operators in \eq{opesix} \cite{Pomarol:2013zra,Falkowski:2014tna}.
Nevertheless, in the presence of a sizeable breaking of the  custodial symmetry, the effects of 
${\cal O}_T$  on  the $Z$ propagator  could partially compensate the $Z\bar\psi\psi$ vertex corrections, in such a way that one linear combination of the operators in \eq{opesix} remains invisible to processes on the $Z$-pole, thus escaping LEP1 constraints~\cite{Gupta:2014rxa}. That particular combination  modifies  $\bar\psi\psi\to W^+W^-$, but the constraints placed by LEP2 data (in the form of the parameter $g_1^Z$ discussed above) are 
 one order of magnitude weaker  \cite{Schael:2013ita,Falkowski:2014tna} than those placed by  $Z$-pole data.   Future measurements at the LHC,
either in $\bar\psi\psi\to V_LV_L$ \cite{other}
or $\bar\psi\psi\to V_Lh$ \cite{Biekoetter:2014jwa},  will instead be able to  provide a more sensitive probe. 
It is amusing that in this case, the best observables to test fermion and Higgs compositeness are those traditionally associated with anomalous TGC~\cite{Hagiwara:1986vm}.

\subsubsection{Fermions as Composite Pseudo-Goldstini}
\label{sec:compoferm2}

In Section~\ref{sec:stronglfermions} we discussed the  scenario
 in which fermions  arise as composite pseudo-Goldstini of a  strong sector with  ${\cal N}>1$ spontaneously broken supersymmetries.
The phenomenological interesting feature of these models is that
the higher-dimensional fermion interactions must involve derivatives, inducing softer effects at low energy.
Indeed, from  \eq{eq:bardeen} we have
\begin{equation}
\delta{\cal A}(\psi\psi\to\psi\psi)\simeq g_*^2\frac{E^4}{m_*^4}\,,
\label{compofermgoldstini}
\end{equation}
and, hence, we expect milder  constraints coming from, e.g., dijet or Drell-Yan collider searches.
 
The above suppression in $\psi\psi\to\psi\psi$ makes  other observables  equally relevant.
 An example is given by the case in which  the Higgs is also composite. 
 Here, beside the effects of Higgs compositeness captured by the SILH scenario, we expect additional genuinely strongly interacting contributions from
 the dimension-8 operators in \eq{goldststructuresN1},  that modify the amplitudes for longitudinal diboson pair production
\begin{equation}\label{eq:softehiggs}
\delta{\cal A}(\bar\psi\psi\to V_LV_L)\simeq g_*^2\frac{E^4}{m_*^4}\,.
\end{equation}
As in the SILH, there are also dimension-6 contributions to 
${\cal A}(\bar\psi\psi\to W_LW_L)$
coming from  ${\cal O}_W$ and ${\cal O}_B$, 
but these are suppressed by weak couplings:
$\delta {\cal A}(\bar\psi\psi\to W_LW_L)\sim g^2E^2/m_*^2$, and  are subleading w.r.t. \eq{eq:softehiggs} for $E\gtrsim (g/g_*)m_*$.
Therefore, in this scenario,  apart from the obvious effects  on Higgs observables
due to the Higgs compositeness,
the most important effects are   modifications in the  high-energy production of dibosons (including neutral $Z_LZ_L$ and $Z_L h$  final states) and quark/lepton pairs, according to \eq{eq:softehiggs} and \eq{compofermgoldstini}  respectively.\\

For the case of composite gauge bosons {\it \`a la}  Remedios 
we also expect the operators in eqs.~(\ref{goldststructuresN1},\ref{bigLag1}) (see also appendix \ref{app:dim8}) to be induced by the strong dynamics.
This leads to 
\begin{equation}\label{dim8remferm}
\delta{\cal A}(\bar\psi\psi\to V_TV_T)\simeq g_*^2\frac{E^4}{m_*^4}\,.
\end{equation}
From vector compositeness \emph{per-se}  we also have  the sizeable dimension-6 operator coefficient $c_{3W}\sim g_*$, implying, from an insertion of an ordinary gauge $\bar\psi\psi V_T$ vertex, a contribution ${\cal A}(\bar\psi\psi\to V_TV_T)\sim gg_*E^2/m_*^2$ (\eq{remffVV}). In this contribution the weakness of the ordinary gauge vertex is compensated by the lower power of $E/m_*<1$.  Nevertheless, the dimension-6 contribution exceeds the SM one, $\sim g^2$, only
for $(g_*/g)E^2/m_*^2\gtrsim 1$, precisely where the dimension-8 contribution \eq{dim8remferm}  becomes fully dominant. In other words, 
 as soon as  contributions  from these scenarios are  larger than the SM, 
 they are dominated by the dimension-8 operators.\footnote{Notice also that in the SM the largest contribution to the $\psi\bar \psi\to WW$ cross-section comes from the $+-$ vector helicity structure, which 
 in the high energy limit  is not modified by dimension-6 effects, but by dimension-8 ones~\cite{other,WWWW}. Therefore, the contribution of dimension-8 versus dimension-6 operators to the cross-section is not only enhanced by the $g_*/g$ factor, but also by this interference term.}

These scenarios are also interesting  because they  give rise to sizeable contributions, \eq{dim8remferm},  to neutral diboson pair production, including photons, without giving rise at the same time to large deviations in electroweak observables or to any other observable affected by dimension-6 operators. 
This   motivates the  possibility to extend the analysis of $q\bar q\to VV$
at the LHC    in a  consistent way,  providing   richer avenues for explorations and prospects.
In particular,  it could be possible  to combine   measurements in $q\bar q\to WW,WZ,W\gamma$
with those with  neutral final-states  $q\bar q\to ZZ,Z\gamma,\gamma\gamma$ 
to test this class of SM deformations~\cite{other}.

\section{Conclusions}
In this paper we have argued that, although the  SM is  weakly-coupled at the electroweak scale,
it may still conceivably emerge from  a strong dynamics just above the TeV scale. The  weakness of  the SM couplings at low-energies 
could   be the consequence  of approximate symmetries: similarly to the pions in QCD, the SM particles could
 correspond to the  lightest  degrees of freedom of the strong sector whose interactions, at very low energy $E<m_*(g/g_*)$,  could be dominated by small symmetry breaking effects, and thus appear weak.

The hierarchy problem
strongly motivates to consider the  Higgs
a composite state directly participating in a strong dynamics above the weak scale.
For this reason composite Higgs   scenarios have been  extensively studied in recent years.
Nevertheless, if a strong sector is supposed to be around the  TeV,
it is also  worth entertaining the possibility that other SM states may be part of it.
The LHC at $13-14$ TeV offers  for the first time  a powerful probe of the nature of the SM particles
in that crucial energy range. The study of the high-energy behaviour $\psi\psi\to \psi\psi$ and $\psi\bar\psi\to VV$ offers at present the best sensitivity, but in the future also the study of $VV\to  VV$ could play a role.

We have here characterized several new patterns of strong dynamics,  encompassing not
just the Higgs boson, but also the SM gauge bosons and  fermions.
The importance of our constructions should be clear. 
They  provide  structurally robust
parametrizations,   based on symmetry and 
dynamics, of  possible SM deformations. In turn each of these parametrizations 
offers a specific ordering of the observables according to their sensitivity to new physics. An overall guideline
for the design of   LHC research strategies is thus devised.
From a more theoretical perspective, the simple and the robust  principles underlying our constructions  provide  well definite power-counting rules for the derivative and field expansions within the effective Lagrangian,
as well as a direct assessment of its domain of applicability. The value of these neat rules is best appreciated 
by considering the situations where they lead to results countering naive expectations. 
To that end, we should stress the recurrent emergence of scenarios where, well within the domain of validity of the effective field theory description, the leading  effects are captured by dimension-8 operators
rather than by dimension-6 ones.

The main new idea considered here has been to have (some of) the SM gauge bosons
arise from a strong sector, in such a way that the the strong dynamics manifests itself purely through higher-derivative
(multipolar) interactions. The extreme IR softness of these strong interactions accounts for their having escaped detection in collider experiments thus far. Their structure  is enforced by a specific symmetry, while the ordinary gauge interactions, which dominate the low energy dynamics, arise as a small deformation of that symmetry. 
In that respect our construction is technically natural. Indeed,
the physical situation depicted by our scenario is fully analogous to the one encountered in  the low-energy effective Lagrangian for neutral atoms and photons, where all interactions are necessarily higher derivatives (multipoles). In that case, the resulting IR softness of the light-matter scattering amplitudes  is the well known cause for the color of the sky. 
Our construction and its consequences, exotic as they may appear, are in fact as natural as that color.

Based on the above idea, we constructed explicit scenarios for new physics, which we dubbed {\it Remedios}, and outlined their phenomenology. Their crucial property is to produce deformations of the SM amplitudes suppressed 
by powers of $E/m_*$ but enhanced by a strong coupling $g_*$: at sufficiently high-energy, but still below the fundamental cut-off scale $m_*$, these deformations can  become sizeable and even
overcome the SM contributions.  
The LHC at $13-14$ TeV is thus well suited to explore these scenarios by looking at  the differential cross-sections for diboson production. Indeed, we have shown that Remedios models always feature large effects in TGC. In particular, they always have a sizable $\lambda_\gamma$, which could be seen in $\psi\bar \psi\to  V_TV_T$. 
As matter of fact, we should stress that we are not aware of other self-consistent and robust constructions where the study of TCG  offers a better option than the search for the underlying new resonances: in order for new physics to show up loud and clear in deformations of the SM amplitudes, below the threshold for new physics, the new dynamics should be rather strong. The strength of the new dynamics  is precisely the novelty offered by our construction. It should be reminded that in the scenarios of composite Higgs the corrections to TCG are not that important, given the SM gauge boson are weakly coupled (elementary) at all energies. Finally, another important class of effects is given by
 deformations   of the vector propagators, in particular those described  by the $W,Y,Z$ parameters of Ref.~\cite{Barbieri:2004qk}, which could be sizable enough to saturate the  experimental bounds coming from LEP. The exploration of these other effects is more for (far) future high-luminosity $e^+e^-$ colliders.

We have also explored the possibility to combine the  Remedios construction for gauge bosons with a composite  PNGB Higgs.
If the latter  arises  from a non-compact coset,
in particular  $ISO(4)/SO(4)$,  additional sizable deviations (beside those expected either in the pure composite  Higgs scenario or in the pure  Remedios scenario) are expected in
$h\to Z\gamma$ and   TGC (in particular, $\kappa_\gamma$), following the  interesting correlation shown in  \eq{corr}.

Finally, we have extended the analysis to  scenarios in which the SM fermions are also strongly coupled at $m_*$.
Four-fermion interactions represent here the main
new physics effect of fermion compositeness,
and constraints from studies of the angular distributions of high-energy dijets
are very strong.
Nevertheless, we have argued that softer fermion interactions (and hence milder constraints) can arise in models based on approximate supersymmetry, where part of, or all of, the SM fermions have the interpretation of pseudo-Goldstini. This idea is not new, as it was already put forward in the 80's \cite{Bardeen:1981df}. However we believe it acquires new vitality offering new perspectives in the general context of strongly coupled scenarios just above the weak scale. For instance,
 the combination of Remedios models with soft composite fermions
implies that the leading effects to $\psi\bar \psi\to  V_TV_T$ come from dimension-8 operators,  which, with respect to the SM,
scale  as $g_*^2 E^4/(g^2m_*^4)$.
A detailed phenomenological analysis of these effects will be given in Ref.~\cite{other}.

Our main results are summarized  in tables~\ref{tab:Silh} and  \ref{operatorsdim6coeff}, where an estimate of the coefficients of the  main induced operators are presented for the different strongly coupled scenarios.
That can be useful to motivate certain searches in Higgs physics, diboson production, or $WW$ scattering
which were not  theoretically well justified before or not properly addressed.
In the end, our constructions offer a map of the possible geography of new physics at the weak scale.
Even if the resulting geography looks exotic, the map is constructed according to well definite structural assumptions. 
It seems to us that, notwithstanding all its limitations,  the charting of new physics that we outlined is conceptually more solid than the standard approach  based on the fully general dimension-6 effective Lagrangian. The occurrence of situations where the dimension-8 operators  dominate  makes that plainly clear.
In the end, the study of the new experimental data will offer a proof or, to the very least, 
a partial disproof of our assumptions.

\subsection*{Acknowledgments}

 We would like to thank  R.~Contino, G.~Dall'Agata, A.~Falkowski, C.~Grojean and A.~Wulzer for useful discussions. The work of AP    has been partly supported by 
the Catalan ICREA Academia Program and  grants
 FPA2014-55613-P, 2014-SGR-1450 and  SO-2012-0234. The work of 
 FR was supported during a good part of this project by the Swiss National Science Foundation, under the Ambizione grant PZ00P2 136932.
 The work of RR is supported by the Swiss National Science Foundation under grants CRSII2-160814 and 200020-150060.

\appendix

\section{Minimal Coupling from Extra-Dimensions}
\label{minimalcoupling}
In this Appendix  we would like to elaborate on  the notion of 
Minimal Coupling (MC), defined
in the main text and in  Ref.~\cite{Giudice:2007fh}, as it  led to some criticism \cite{Jenkins:2013fya}.
We will do so by  illustrating how the arguably simplest 5D construction
leads to a low-energy effective Lagrangian  effectively satisfying MC.
Let us consider, for the sake of the argument, a 5D Yang-Mills theory compactified on a circle  $S_1$ of radius $R$. At the tree level the light degrees of freedom are represented by the vector zero modes $A_\mu^a$ and by the Wilson lines $W^a\equiv \frac{1}{2\pi R}\oint A_5^a$. The low-energy effective Lagrangian is determined by the parameters of the microscopic 5D Lagrangian describing physics at distances $\ll R$, and by the effects at the KK threshold.
The simplest option  is to assume  only  one microscopic scale at which the 5D theory is strongly coupled. Equivalently, that amounts to assuming the coefficients of the 5D Lagrangian satisfy naive dimensional analysis (NDA). Given the 5D coupling $g_5$, a simple analysis of loop effects allows to conclude that, in the absence of intervening new physics at lower scales, the theory becomes strong at around the scale
\be
\Lambda_5\equiv  \frac{16\pi^2}{g_5^2}\, ,
\ee
and therefore using the NDA ansatz:
\be
{\cal L}_5= \frac{\Lambda_5}{16\pi^2} L\left (D_M/\Lambda_5\right )=\frac{\Lambda_5}{16\pi^2}\left \{ -\frac{1}{4}F_{MN}^2+ \frac{c_1}{\Lambda_5^2}F_{MN} D_P^2 F^{MN}+\frac{c_2}{\Lambda_5^2}F_{MN}^3+\dots\right \}\,.
\label{nda}
\ee
Notice that the NDA ansatz is compatible with naturalness, in that, starting from the leading $F_{MN}^2$ term, all others would be generated from loops with coefficients of precisely the NDA size. For instance, the terms associated with $c_{1,2}$ are expected to arise from log divergences at 2-loops. One could even slightly generalize the NDA assumption by assuming that the 5D physics is itself characterized by one scale $M_5$ and one dimensionless coupling $\tilde g_{*}$:
\be
{\cal L}_5= \frac{M_5}{\tilde g_*^2} L\left (D_M/M_5\right )\,,
\label{nda*}
\ee
and with the identification
\be
\frac{M_5}{\tilde g_*^2}\equiv \frac{1}{g_5^2}\qquad\Longrightarrow \qquad M_5=\Lambda_5\frac{\tilde g_*^2}{16\pi^2}<\Lambda_5\,.
\ee
The scale $M_5$ could be interpreted as the scale where new 5D resonances are expected. The NDA case \eq{nda} simply corresponds to the strongly coupled limit $\tilde g_*\sim 4\pi$. Considering now the general ansatz in \eq{nda*}, it is easy to  deduce the general structure of the effective Lagrangian below the compactification scale. It will be organized as a triple expansion in loops 
 and in inverse powers of $ m_{KK}$ and  $M_5$. The loop expansion parameter $g_*$ and resonance scale $m_*$ are identified according to eq.~(\ref{KKg_5}), but here, for a more precise identification, we take $m_*=1/(\pi R)$,
 as the physical scale of a compactified extra dimension is determined by its length $\pi R$.
  The expansion in $1/M_5$ can be expressed in terms of the dimensionless parameter $m_*/M_5\ll1$. The result is then
 \be
 {\cal L}_{eff}=\frac{m_*^4}{g_*^2} \sum_{n,m}\left (\frac{g_*^2}{16\pi^2}\right )^n\left (\frac{m_*^2}{M_5^2}\right )^mL_{n,m}\left(\frac{D_\mu}{m_*}, \frac{F^a_{\mu\nu}}{m_*},\frac{W^a}{m_*}\right)\,.
 \ee
 The leading order term $L_{0,0}$ simply arises by integrating out the KK resonances at tree level in a 5D theory with the minimal Yang-Mills Lagrangian, that is the first term in eq.~(\ref{nda}). This Lagrangian enjoys some accidental properties, which went by the name {\it {minimal coupling}} in Ref.~\cite{Giudice:2007fh}. In particular, operators in the class of ${\cal O}_{HW}$, ${\cal O}_{BB}$ and ${\cal O}_{3W}$ do not appear in $L_{0,0}$. These terms however appear in general at the next order in the expansion and are thus suppressed by either a loop factor or by
 \be
 \frac{m_*^2}{M_5^2}\sim \left (\frac{ g_*^2}{\tilde g_*^2}\right )^2\,.
 \ee
Notice that in the case of a 5D theory based on the ordinary NDA, we have $\tilde g_*\sim 4\pi$, so that the above factor is equivalent to a 2-loop suppression. From our construction it is clear that the only way to fully eliminate the suppression dictated by MC  is to assume $M_5\sim m_*$, corresponding to the full lack of validity of the 5D description.

The above arguments indicate that a version of MC necessarily applies in theories that admit a range of lengths where they are described by  a weakly coupled 5D theory. We cannot conclude that the same pattern of suppression must appear in other scenarios, like large-$N$ gauge theories or in the effective descriptions of strings. Yet there are intriguing indications that a suppression of the same type also exists in those other contexts. But we do not have a proof that must happen by necessity.

 \section{An Accidentally Light Higgs}\label{sec:ALH}

Here we shall illustrate the scenario where the Higgs boson, while arising from a strong dynamics, cannot be interpreted as a PNGB. 
Its small mass purely appears as the result of some unexplained tuning. We dub this scenario
the Accidentally Light Higgs (ALH).
For the  ALH the  Higgs potential is not dictated by selection rules, and 
 one  expects the  generic function (working with the neutral Higgs component $h$):
\be\label{ALHpot}
V(h)= \frac{m_*^4}{g_*^2}F\left(\frac{h^2}{f^2}\right)={m_*^2}{f^2}\left \{a_2\frac{h^2}{f^2}+a_4\left(\frac{h^2}{f^2}\right )^2+a_6\left(\frac{h^2}{f^2}\right )^3+\dots\right \}\,.
\ee 
To  reproduce a VEV $\langle h\rangle =v$ and a mass $m_h$ much below
their natural expectations,  $f\equiv m_*/g_*$ and  $m_*\sim$ TeV respectively,
one needs an (accidental)  tuning of the parameters of the potential, away from the generic expectation $a_i\sim O(1)$. 
There are virtually as many ways to tune as there are parameters in the potential, that is infinitely many.  Considering the Higgs VEV and mass,  
which are determined  by  ($\xi \equiv v^2/f^2$)
\bea
 \xi\,\,\Longleftrightarrow\,\, F'(\xi)&=&a_2+2a_4\xi+3a_6\xi^2+\dots=0\\
\frac{m_h^2}{m_*^2}=2 F''(\xi)\xi&=&2(2a_4+6a_6\xi+\dots)\xi \, ,
\eea
one can however qualitatively distinguish three main regions of parameter space, according to whether $\xi (m_*/m_h)$ is smaller than, comparable to,  or larger than $O(1)$. The  first region, $\xi m_h/m_*\ll 1$, can also be characterized by
 $a_2\ll a_4^2\ll a_{n\geq 6}\sim O(1)$. In this region, the non-renormalizable terms ($a_{n\geq 6}$) are just a small perturbation around the minimum controlled by the first two terms. 
 On can for instance check that the relative size of the deviations from the SM in the Higgs self-couplings is controlled by $\sim a_2/a_4^2 \sim \xi m_*/m_h\ll 1$. The smallness of these corrections is controlled by the smallness of $\xi$, which in turns follows from
the  important tuning  $a_2\ll a_4^2\ll 1$. A milder tuning is achieved for  $a_2\sim a_4^2\ll a_{n\geq 6}\sim O(1)$ which also generically corresponds to the   intermediate region $\xi   m_*/m_h\sim O(1)$. Here we  have 
   $O(1)$ deviations from the SM in the Higgs self-couplings, even in the presence of a separation of scales. The reason for such a seemingly non-decoupling effect is the sizable coupling $g_*^2\gg \lambda_h\sim a_4 g_*^2$ controlling the higher order terms, and the culprit is just the tuning. 
Finally, in the third region $\xi m_h/m_*\gg 1$, while remaining at $\xi \ll 1$,  the  deviations from the SM  in Higgs self-interactions can be  larger than $ O(1)$. That result is nicely illustrated by focussing on
 the trilinear self-coupling which,  from \eq{ALHpot}, reads
\be
\lambda_{3h}=\frac{6m_h^2}{v}\left (1+\frac{2}{3} \frac{F'''(\xi)\xi}{F''(\xi)}\right )\,,
\ee
 where we singled out the SM result, corresponding to $F'''=0$. The region $\xi m_h/m_*\gg 1$ corresponds to a situation  where $a_{n\geq 6}$, and not $a_2$,
 are used to tune 
 $F''$ at the minimum, and thus $m_h^2$,  very small in such a way that $F'''\xi/F''\gg 1$. 
This third region is thus characterized by an additional tuning of the physical Higgs mass $m_h$. An explicit example is obtained for instance by considering small perturbations around the tuned potential $F(\xi)\equiv (\xi-\xi_0)^3$ for which $m_h=0$. Notice indeed that we can write the relative correction to the trilinear as 
 \be
 F'''(\xi)\xi/F''(\xi)\sim (\xi m_*/m_h)^2 F'''(\xi)\sim (\xi m_*/m_h)^2\,.
 \ee
showing that,  in the absence of further cancellations in $F'''$, it is precisely controlled by  $\xi m_*/m_h$.

When  $\xi m_*/m_h\gg 1$ the Higgs self-coupling can in principle be as large as $\sim m_h g_*$, which implies a large scattering amplitude for $hh\to hh$:   
\be
{\cal A}(hh\to hh)\sim g_*^2\,,
\ee
already at an energy of order $m_h$. For large $g_*\sim 4\pi$, Higgs self-interactions could be rather strongly coupled just around threshold. This scenario is already  constrained by the LHC data on double Higgs production, but, to our knowledge, a detailed study is missing.

\section{Dimension-8 Operators}\label{app:dim8}

In this Appendix we list the  CP-even, custodial preserving, dimension-8 operators that can give  important $\sim g_*^2/m_*^4$  contact-interaction
contributions to $2\to 2$ scatterings at high-energy.
We limit ourselves to processes that involve  at least a pair of bosons and use field redefinitions (equivalent to equations of motion) to rewrite terms with  derivatives (e.g. $D_\mu B^{\mu\nu}$, $D_\mu W^{a\mu\nu}$, $\Box H$, $\dslash\psi$) as terms with  fields  (see also \cite{Henning:2015daa}). Operators of the form $|H|^2{\cal O}_6$, with ${\cal O}_6$ a dim-6 operator, can be  read directly from, e.g., Ref.~\cite{Grzadkowski:2010es,Elias-Miro:2013mua} and generalization  to operators with gluons is straightforward, so we omit them here. 

\vspace{1mm}
\noindent
\framebox[1.5cm][l]{$(X_{\mu\nu})^4$}
In models with the \emph{Remedios} structure, we find
\begin{align}
SU(2)_L:\quad\quad&_8{\cal O}_{4W}= W^a_{\mu\nu}  W^{a\, \mu\nu} W^b_{\rho\sigma}  W^{b\, \rho\sigma}&\quad\quad
_8{\cal O}^\prime_{4W}= W^a_{\mu\nu}  W^{b\, \mu\nu} W^a_{\rho\sigma}  W^{b\, \rho\sigma}\\
&_8{\cal O}_{4\widetilde W}= W^a_{\mu\nu}  W^{a\, \nu\rho} W^b_{\rho\sigma}  W^{b\, \sigma\mu}&\quad\quad
_8{\cal O}^\prime_{4\widetilde W}= W^a_{\mu\nu}  W^{b\, \nu\rho} W^a_{\rho\sigma}  W^{b\, \sigma\mu}\\\nn\\
U(1)_Y:\quad\quad&_8{\cal O}_{4B}= B_{\mu\nu}  B^{\mu\nu} B_{\rho\sigma}  B^{\rho\sigma}&\quad\quad
_8{\cal O}_{4\widetilde B}= B_{\mu\nu}  B^{\nu\rho} B_{\rho\sigma}  B^{\sigma\mu}\\\nn\\
SU(2)_L\times U(1)_Y:\quad\quad&
_8{\cal O}_{2WB}= W^a_{\mu\nu}  W^{a\, \mu\nu} B_{\rho\sigma}  B^{\rho\sigma}&\quad\quad
_8{\cal O}^\prime_{2WB}= W^a_{\mu\nu}  B^{\mu\nu} W^a_{\rho\sigma}  B^{ \rho\sigma}\\
&_8{\cal O}_{2\widetilde W\widetilde B}= W^a_{\mu\nu}  W^{a\, \nu\rho} B_{\rho\sigma}  B^{\sigma\mu}&\quad\quad
_8{\cal O}^\prime_{2\widetilde W\widetilde B}= W^a_{\mu\nu}  B^{\nu\rho} W^a_{\rho\sigma}  B^{ \sigma\mu}\,.
\end{align}
Notice that   $B_{\mu\nu}  \widetilde B^{\mu\nu} B_{\rho\sigma}  \widetilde B^{\rho\sigma}$ (and similar for $W$) can be eliminated in favor of the above using the properties of the Levi-Civita tensor.

\noindent
\framebox[2.3cm][l]{$D\psi^2(X_{\mu\nu})^2$} Strongly interacting fermions and vectors generate
\begin{gather}
_8{\cal O}_{TWW}={\cal T}^{\mu\nu} W_{\mu\rho}^aW^{a\,\rho}_\nu\quad\quad 
 _8{\cal O}_{TBB}= {\cal T}^{\mu\nu} B_{\mu\rho}B^{\rho}_\nu\label{eqTvv}\\
 _8{\cal O}_{TWB}= {\cal T}^{a\,\mu\nu}W_{\mu\rho}^a B^{\rho}_\nu
 \label{ftt2}
\end{gather}
where ${\cal T}^{\mu\nu}=\frac{i}{4}\bar\psi(\gamma^\mu\lra {D^\nu}+\gamma^\nu \lra {D ^\mu})\psi$ and ${\cal T}^{a,\,\mu\nu}=\frac{i}{4}\bar\psi(\gamma^\mu\lra {D^\nu}+\gamma^\nu \lra {D ^\mu})\sigma^a\psi$ for $SU(2)_L$ doublets.  On the other hand
$
_8{\cal O}_{JWW}=\epsilon^{abc} J_\psi^{a\nu}W_{\rho\mu}^b \lra {D_\nu} \widetilde W^{c\,\rho\mu}$, $
_8{\cal O}_{JWB}=  J_\psi ^{a\nu}W_{\rho\mu}^a \lra {D_\nu} \widetilde B^{\rho\mu}$
are odd under both $C$ and $P$, and CP even ($J_\psi ^{a\nu}=\bar\psi\gamma^\nu\sigma^a\psi$, $J_\psi ^{\nu}=\bar\psi\gamma^\nu\psi$ denote universal $SU(2)_L\times U(1)_Y$ currents, the extension to other cases being straightforward).  Operators of the form $ J_\psi^\nu B^{\mu\rho}D_\mu \widetilde B_{\rho\nu}$ (and similarly for $W^a_{\mu\nu}$), or operators involving $\bar\psi(\gamma^\mu\lra {D^\nu}-\gamma^\nu \lra {D ^\mu})\psi$ vanish due to Bianchi identities.
The operators $_8{\cal O}_{JWW}$, $_8{\cal O}_{JWB}$ and $ _8{\cal O}_{TWB}$  cannot arise in the model of 
Section~\ref{so5remedios}, as they are not singlets under $\cal G$ in \eq{so5su2}  - the former are also suppressed for $\psi$ pseudo-Goldstini. 

\vspace{1mm}
\noindent
\framebox[1.3cm][l]{$D^4H^4$} In models where  the Higgs is composite,
\begin{equation}\label{dim8hhscatt}
_8{\cal O}_{\{D\}H}=(D_{\{\mu} H^\dagger D_{\nu\}}H )^2\quad\quad_8{\cal O}_{DH}=(D_\mu H^\dagger D^\mu H)^2
\end{equation}
mediate interaction between four (longitudinal) vectors, that might be relevant in the model $ISO(4)/SO(4)$ of Section \ref{iso}, where the leading contribution from ${\cal O}_H$ is suppressed. 
Operators that involve the $\mu\leftrightarrow\nu$ (anti)symmetric part of $D_\mu H^\dagger D^\nu H$ ($D_\mu H^\dagger \sigma ^aD^\nu H$), transform as a ${\bf(1_L,3_R)}$ and ${\bf(3_L,3_R)}$ of $SU(2)_L\times SU(2)_R$ and break custodial symmetry; while the custodial preserving $(D_{[\mu} H^\dagger\sigma^a D_{\nu]}H )^2$  can be rewritten as \eq{dim8hhscatt} using the properties of Pauli matrices.  
For completeness, we list
$\left( H^\dagger D_\mu D_\nu H  + D_\mu D_\nu H^\dagger H \right)^2$ and
$\left( H^\dagger \sigma^ aD_\mu D_\nu H  - D_\mu D_\nu H^\dagger\sigma^ a H \right)^2$
 which however vanish in $ISO$ models.

\vspace{1mm}
\noindent
\framebox[2.5cm][l]{$D^2H^2(X_{\mu\nu})^2$} On the other hand,
\begin{gather}\label{HHWW}
_8{\cal O}_{HWW}= D_\mu H^\dagger D_\nu H\, W^{a\,\mu}_{\rho}  W^{a \nu\rho}\,,\quad _8{\cal O}_{HBB}= D_\mu H^\dagger D_\nu H\, B^\mu_{\rho}  B^{\nu\rho}\\
_8{\cal O}^\prime_{HWW}= D_\mu H^\dagger\sigma^a D_\nu H\, W^{b\,\mu}_{\rho}  W^{c \nu\rho}\epsilon^{abc}\,,\quad _8{\cal O}_{HWB}=D_\mu H^\dagger\sigma^a D_\nu H\, W^{a\,\mu}_{\rho}  B^{\nu\rho}\label{HHWW2}
\end{gather}
contribute to processes with two transverse and two longitudinal modes, although \eq{HHWW2} are forbidden in the model of Section~\ref{so5remedios} because they break the global symmetry $\cal G$, but are allowed if the Higgs originates from the $ISO(4)/SO(4)$ coset.
The structure 
$( H^\dagger \sigma^ aD_\mu D_\nu H$ $  - D_\mu D_\nu H^\dagger\sigma^ a H )W^a_{\mu\rho}  B^{\nu\rho}$ is instead forbidden in both models as it breaks both  $\cal G$ and $ISO$.
  
\vspace{1mm}
\noindent
\framebox[1.7cm][l]{$D^3H^2\psi^2$} If the fermions are pseudo-Goldstini, 
\bea\label{eqTH}
_8{\cal O}_{TH}&=&{\cal T}^{\mu\nu} D_\mu H^\dagger D_\nu H
\eea
mediates the leading interaction between fermions and two  longitudinal gauge bosons, including effects in $Z_LZ_L$.
In non-supersymmetric models, the structure 
$_8{\cal O}^\prime_{JH}=J^{a\,\nu} D^\mu H^\dagger\sigma^a \lra D_\nu D_\mu H
$
also arises, but is of limited interest as it clearly only contributes to $W_L^+W_L^-$ production and is always subdominant w.r.t. dimension-6 effects from ${\cal O}^\psi_{L,R}$~and~${\cal O}^{(3)\psi}_{L}$. 
 
 \vspace{1mm}
\noindent
\framebox[2.2cm][l]{$DH^2\psi^2X_{\mu\nu}$} Finally, for completeness, we mention contributions to $\bar q q\to V_T Z^\prime_L$ when fermions and Higgs are composite and gauge bosons are dipole-strong. The operator $J^\mu_H J_\psi^\nu B_{\mu\nu}$ (with $J^\mu_H= H^\dagger \lra {D^\mu} H$) is forbidden by custodial symmetry. Instead $J^\mu_H J_\psi^{a\,\nu} W^a_{\mu\nu}$, $J^{a\,\mu}_H J_\psi^{\nu} W^a_{\mu\nu}$ and
$J^{a\,\mu}_H J_\psi^{b\nu} W^c_{\mu\nu}\epsilon^{abc}$ are suppressed in the $SO(5)/SO(4)$ model as they break the global symmetry and in $ISO$ models as they break the shift symmetry.

\end{document}